\documentclass[aps,groupedaddress,floatfix,nofootinbib,preprintnumbers,eqsecnum,twocolumn]{revtex4}
\usepackage{amssymb,amsmath,graphicx,multirow}

\begin{document}

\preprint{UH511-1212-2013}

\title{Dynamical Dark Matter and the Positron Excess in Light of AMS\\}
\author{Keith R. Dienes$^{1,2,3}$\footnote{E-mail address:  {\tt dienes@physics.arizona.edu}},
      Jason Kumar$^{4}$\footnote{E-mail address:  {\tt jkumar@phys.hawaii.edu}},  
      Brooks Thomas$^{4}$\footnote{E-mail address:  {\tt thomasbd@phys.hawaii.edu}}}
\affiliation{
     $^1$ Physics Division, National Science Foundation, Arlington, VA  22230  USA\\
     $^2$ Department of Physics, University of Maryland, College Park, MD  20742  USA\\
     $^3$ Department of Physics, University of Arizona, Tucson, AZ  85721  USA\\
     $^4$ Department of Physics, University of Hawaii, Honolulu, HI 96822  USA}

\begin{abstract}
The \mbox{AMS-02} experiment has recently released data which confirms a rise in the 
cosmic-ray positron fraction as a function of energy up to approximately $350$~GeV.~ 
Over the past decade, attempts to interpret this positron excess in terms of 
dark-matter decays have become increasingly complex and have led to a number of 
general expectations about the decaying dark-matter particles:  such particles cannot 
undergo simple two-body decays to leptons, for example, and they must have rather 
heavy TeV-scale masses.  In this paper, by contrast, we show that Dynamical Dark 
Matter (DDM) can not only match existing \mbox{AMS-02} data on the positron excess, but also 
accomplish this feat with significantly lighter dark-matter constituents undergoing 
simple two-body decays to leptons.  Moreover, we demonstrate that this can be done 
without running afoul of numerous other competing constraints from FERMI and Planck 
on decaying dark matter.  Finally, we demonstrate that the Dynamical Dark Matter 
framework makes a fairly robust prediction that the positron fraction should level 
off and then remain roughly constant out to approximately 1~TeV, without experiencing 
any sharp downturns.  Indeed, if we interpret the positron excess in terms of decaying 
dark matter, we find that the existence of a plateau in the positron fraction at 
energies less than 1~TeV may be taken as a ``smoking gun'' of Dynamical Dark Matter.     
\end{abstract}

\maketitle

\newcommand{\newc}{\newcommand}
\newc{\gsim}{\lower.7ex\hbox{$\;\stackrel{\textstyle>}{\sim}\;$}}
\newc{\lsim}{\lower.7ex\hbox{$\;\stackrel{\textstyle<}{\sim}\;$}}
\makeatletter
\newcommand{\biggg}{\bBigg@{3}}
\newcommand{\Biggg}{\bBigg@{4}}
\makeatother

\def\vac#1{{\bf \{{#1}\}}}

\def\beq{\begin{equation}}
\def\eeq{\end{equation}}
\def\beqn{\begin{eqnarray}}
\def\eeqn{\end{eqnarray}}
\def\calM{{\cal M}}
\def\calV{{\cal V}}
\def\calF{{\cal F}}
\def\half{{\textstyle{1\over 2}}}
\def\quarter{{\textstyle{1\over 4}}}
\def\ie{{\it i.e.}\/}
\def\eg{{\it e.g.}\/}
\def\etc{{\it etc}.\/}


\def\inbar{\,\vrule height1.5ex width.4pt depth0pt}
\def\IR{\relax{\rm I\kern-.18em R}}
 \font\cmss=cmss10 \font\cmsss=cmss10 at 7pt
\def\IQ{\relax{\rm I\kern-.18em Q}}
\def\IZ{\relax\ifmmode\mathchoice
 {\hbox{\cmss Z\kern-.4em Z}}{\hbox{\cmss Z\kern-.4em Z}}
 {\lower.9pt\hbox{\cmsss Z\kern-.4em Z}}
 {\lower1.2pt\hbox{\cmsss Z\kern-.4em Z}}\else{\cmss Z\kern-.4em Z}\fi}
\def\OmegaDM{\Omega_{\mathrm{CDM}}}
\def\Omegatot{\Omega_{\mathrm{tot}}}
\def\rhocrit{\rho_{\mathrm{crit}}}
\def\arcsinh{\mbox{arcsinh}}
\def\BRgamma{\mathrm{BR}_{\lambda}^{(2\gamma)}}
\def\OmegaDM{\Omega_{\mathrm{CDM}}}
\def\tnow{t_{\mathrm{now}}}
\def\Omegatotnow{\Omega_{\mathrm{tot}}^\ast}
\def\erf{\mathrm{erf}}
\def\rhototloc{\rho^{\mathrm{loc}}_{\mathrm{tot}}}
\def\Ecut{E_{\mathrm{cut}}}
\def\Emax{E_{\mathrm{max}}}


\input epsf





\section{Introduction\label{sec:intro}}


One of the most urgent problems facing particle physics, astrophysics, and cosmology
today is that of understanding the nature of dark matter.  Fortunately, a 
confluence of emerging data from direct-detection, indirect-detection, 
and collider experiments suggests that major progress may soon be at hand.  A 
potentially important ingredient in this mix may involve recent 
results~\cite{AMS02DataNew} from the \mbox{AMS-02} experiment on the flux of cosmic-ray 
positrons at energies up to $350$~GeV.~  These results confirm the 
anomalous and puzzling results observed by earlier cosmic-ray detectors such as 
HEAT~\cite{HEATData1997PosFrac}, AMS-01~\cite{AMS01Data2002,AMS01Data2007}, 
PAMELA~\cite{PAMELAData2008,PAMELAData2010PosFrac}, 
and FERMI~\cite{FERMIPositronData2011} which indicate
that the positron fraction --- \ie, the ratio of the 
differential flux $\Phi_{e^+}$ of cosmic-ray positrons to the 
combined differential flux $\Phi_{e^-} + \Phi_{e^+}$ 
of cosmic-ray electrons and positrons --- actually rises as a function
of particle energy $E_e$ for energies $E_e \gtrsim 10$~GeV.~  
Since the positron fraction is generally expected to fall with energy in
this energy range,
the observed positron excess suggests that some unanticipated 
physics might be in play.  While many possibilities exist, one natural idea is 
that these positrons may be produced via the annihilation or decay 
of dark-matter particles within the galactic halo.
Unfortunately, this rise in the positron flux occurs without any other distinctive features,
and no downturn at high energies --- a standard prediction of
the most straightforward dark-matter models --- is apparent. 
This rise in the positron fraction therefore poses a major challenge for
any potential interpretation in terms of dark-matter physics.

At first glance, it might seem relatively straightforward to interpret
the observed positron excess in terms of annihilating or decaying dark-matter 
particles within the galactic halo. 
However, such a dark-matter interpretation 
of the cosmic-ray positron excess is tightly constrained by a number of 
additional considerations.
For example, no corresponding excess is observed in the flux of 
cosmic-ray antiprotons~\cite{PAMELAAntiproton}, a fact which 
significantly constrains the particle-physics properties 
of possible dark-matter candidates.  Indeed, these constraints are 
particularly severe for dark-matter candidates which annihilate or 
decay either predominately to strongly-interacting Standard-Model (SM) 
particles or to particles such as $W^\pm$ or $Z$ whose subsequent decays 
produce such particles with significant frequency.  
For this reason, the most natural dark-matter candidates 
which can explain the observed positron excess are those  
which annihilate or decay primarily to charged leptons.

However, even such ``leptophilic'' dark-matter candidates are significantly 
constrained by cosmic-ray data.
Precise measurements of the combined flux spectrum of cosmic-ray electrons and 
positrons by the FERMI collaboration~\cite{FERMIPositronData2010} further
restrict the range of viable dark-matter models of the observed positron excess.
Moreover, constraints on the production of photons are also quite stringent.
For example, high-energy photons produced by a cosmological population of dark-matter
particles contribute to the diffuse extragalactic gamma-ray background which
has been measured by the FERMI collaboration~\cite{FERMIGammaRays}.  
In addition, the energy released via the annihilation or decay 
of dark-matter particles in the early universe can also lead to a 
reionization of the thermal plasma at or after the time of last scattering.
This in turn induces a modification of the observed temperature and polarization 
fluctuations of the cosmic microwave background (CMB).~
As a result, CMB data from BOOMERANG~\cite{BOOMERANG}, ACBAR~\cite{ACBAR}, 
WMAP~\cite{WMAP}, and Planck~\cite{PlanckResults} significantly 
constrain the rate at which dark matter can annihilate or decay during and 
after the recombination epoch. 

A number of scenarios have been advanced over the past decade to reconcile the 
observed positron fraction with these additional constraints~\cite{block}.  As 
these constraints have sharpened over time, the corresponding dark-matter models 
have also grown in complexity and sophistication --- a trend which has only
continued~\cite{newAMSblock} since the release of the most recent \mbox{AMS-02} data.
For example, one current 
possibility~\cite{HooperAMS} involves a dark-matter particle which annihilates 
or decays into exotic intermediate states which only subsequently decay 
into $\mu^\pm$ or $\pi^\pm$.   Other 
possibilities~\cite{IbarraThreeBody,CosmicThreeBodyRayDM} involve dark-matter 
particles which decay primarily via three-body processes of the form 
$\chi \rightarrow \psi\, \ell^+\ell^-$, where $\psi$ is an additional, 
lighter dark-sector field and where
$\ell^\pm =\{e^\pm,\mu^\pm,\tau^\pm\}$.  There also exist other three-body-decay 
models~\cite{BargerThreeBodyPAMELA,IbeSUSY3BodyAMS} in which all of the final 
states are in the visible sector.  In each case, these features are required 
in order to ensure that the resulting electron and positron flux spectra are 
significantly ``softer'' (\ie, broader, more gently sloped) than those produced
by a dark-matter particle undergoing a two-body decay directly to SM states.
Indeed, only such softer spectra can provide a satisfactory combined fit to the observed 
positron fraction and to the $\Phi_{e^+} + \Phi_{e^-}$ flux spectrum observed by 
FERMI while simultaneously satisfying all other phenomenological constraints.

Taken together, these attempts have led to certain expectations concerning the nature
of the dark-matter particle whose decays or annihilations might explain the positron excess.
Specifically, it is expected that this dark-matter particle will not undergo two-body
decays to Standard-Model states, but will instead decay through more complex   
decay patterns such as those involving non-trivial intermediate states or
three-body final states. 
As discussed above, this is necessary in order to soften the kinematic spectrum 
associated with such single-particle dark-matter candidates.
Second, it is also expected that such dark-matter particles must be relatively heavy,
with masses $\sim {\cal O}$(TeV), in order to properly explain the measured
positron excess.  This is unfortunate, since leptophilic particles with such
heavy masses are typically difficult to probe via other experiments
(\eg, existing collider experiments) which 
provide complementary probes of the dark sector.
Finally, we note that all current dark-matter-based attempts at explaining the 
observed positron excess inevitably predict that the positron fraction will
experience a relatively sharp downturn at energies 
which do not greatly exceed current sensitivities.
Indeed, a relatively sharp downturn is in some sense required by the decay 
kinematics of such dark-matter candidates.

In this paper, we will show that Dynamical Dark Matter 
(DDM)~\cite{DynamicalDM1,DynamicalDM2} can provide an entirely different perspective 
on these issues.  First, we shall demonstrate that a leptonically-decaying DDM ensemble 
can successfully account for the observed positron excess and combined cosmic-ray $e^\pm$ 
flux without running afoul of any other applicable constraints on decaying or annihilating 
dark-matter particles.  Second, we shall show that DDM can do this entirely with dark-matter components 
undergoing simple two-body decays to leptons --- indeed, more complicated decay 
phenomenologies are not required.  Third, we shall find that the DDM components which 
play the dominant role in explaining the positron excess are themselves relatively 
light, with masses only in the ${\cal O}(200-500)$~GeV range.  This is an important
distinction relative to more traditional models, opening up the possibility of 
correlating these positron-flux signatures with possible missing-energy signatures 
in collider experiments.  This would then allow a more tightly constrained, 
complementary approach to studying such dark-matter candidates.
Indeed, as we shall see, DDM accomplishes all of these feats by providing
an {\it alternative}\/ method of softening the flux spectra --- not 
through a complicated set of dark-matter decay/annihilation channels (and thus
complicated particle kinematics), but instead through a richer and more complex 
dark sector itself.

But perhaps most importantly, we shall show that DDM also makes a fairly
firm prediction for the positron fraction at energies {\it beyond}\/ 350~GeV:
the positron fraction will level off and remain roughly 
constant all the way up to energies of approximately $1$~TeV.~
Indeed, as we shall find, this behavior for the positron fraction emerges 
for most of the viable regions of DDM parameter space.
Hence, within such regions, the DDM framework predicts that no abrupt downturn in the
positron fraction will be seen.  This is a marked difference relative
to most traditional dark-matter models which seek to explain
the positron excess:  indeed, most of these models predict either a continuing rise
in the positron fraction or the onset of a downturn, but cannot easily accommodate
a relatively flat plateau.  Thus, if we interpret the positron excess seen by \mbox{AMS-02} 
as resulting from dark-matter annihilations or decays, a relatively flat plateau in the 
positron fraction at energies less than 1~TeV may be taken as a ``smoking gun'' of 
Dynamical Dark Matter.
 
This paper is organized as follows.  In Sect.~\ref{sec:ensembles}, we briefly 
review the general properties of DDM ensembles and introduce the general 
parametrizations we shall use in order to characterize these ensembles in our analysis.  
In Sect.~\ref{sec:propagation}, we then discuss the $e^\pm$ injection spectra produced 
by the decays of a DDM ensemble and show how these spectra are modified upon
propagation through the interstellar medium.  In Sect.~\ref{sec:constraints}, we then 
discuss the additional considerations which further constrain decaying dark-matter candidates
and examine how these considerations apply to DDM ensembles.  
Our main results appear in Sect.~\ref{sec:results}, where we demonstrate that DDM 
ensembles can indeed reproduce the observed positron excess while simultaneously satisfying
all relevant constraints --- even with relatively light DDM constituents undergoing simple
two-body leptonic decays.  We also demonstrate that most of the viable DDM parameter space
leads to the prediction of a positron fraction which levels off and remains roughly constant 
out to energies of approximately $1$~TeV.~ Even though (as we shall see) there exist other regions 
of viable DDM parameter space for which the predicted positron excess can experience a downturn 
(or even an oscillation!)\ as a function of energy, we shall explain why we nevertheless believe 
that the existence of a plateau in the positron fraction can serve as a ``smoking gun'' for the 
Dynamical Dark Matter framework as a whole.  
In Section~\ref{sec:varyinginputs}, we then
discuss the extent to which these results continue to apply when our fundamental theoretical assumptions
and computational procedures are altered.
Finally, in Sect.~\ref{sec:conclusions}, we conclude 
with a summary of our results and a discussion of their implications for distinguishing between decaying 
DDM ensembles and other proposed explanations for the positron excess, including those
involving purely traditional astrophysical sources.


\section{The DDM Ensemble:  ~Fundamental Characteristics\label{sec:ensembles}}


Dynamical Dark Matter (DDM)~\cite{DynamicalDM1,DynamicalDM2} is an alternative 
framework for dark-matter physics in which the requirement of dark-matter stability
is replaced by a balancing of lifetimes against cosmological abundances across an 
ensemble of individual dark-matter components with different masses, lifetimes, and
abundances.  It is this DDM ensemble which collectively serves as the dark-matter 
``candidate'' in the DDM framework, and which collectively carries the observed 
dark-matter abundance $\Omega_{\rm CDM}$.  Likewise, it is the balancing between 
lifetimes and abundances across the ensemble as a whole which ensures the phenomenological 
viability of the DDM framework~\cite{DynamicalDM2,DynamicalDM3}.  In some sense the DDM 
ensemble is the most general dark sector that can be envisioned, reducing to a standard 
stable dark-matter candidate in the limit that the number of dark-matter components is 
taken to one.  However, in all other cases, stability is not an absolute requirement
in the DDM framework (a feature which distinguishes DDM from other multi-component 
dark-matter scenarios), but instead depends, component by component, on the corresponding 
cosmological abundances.  As has been discussed in 
Refs.~\cite{DynamicalDM1,DynamicalDM2,DDMColliders}, DDM ensembles appear naturally in 
many extensions to the Standard Model, including string theory and theories with large 
extra spacetime dimensions, and not only possess a highly non-trivial cosmology but can 
also lead to many striking signatures at colliders~\cite{DDMColliders} and
direct-detection experiments~\cite{DDMDirectDet} --- signatures which transcend those 
associated with traditional dark-matter candidates.
Indeed, DDM ensembles are fairly ubiquitous,
and can also potentially arise in a variety of
additional contexts ranging from
theories such as the axiverse~\cite{axiverse} to
theories involving large hidden-sector gauge groups
and even theories exhibiting warped stringy throats~\cite{Mazumdar}.

It is the purpose of this paper to examine the behavior of the positron flux within the 
context of the general DDM framework, and thereby study the implications of the DDM 
framework for {\it indirect}\/-detection experiments.  Because the DDM framework lacks 
dark-matter stability as a founding principle, discussions of decaying dark matter
(such as those possibly leading to a cosmic-ray positron excess) are particularly relevant 
for DDM.~  Indeed, one important characteristic of DDM is that the DDM dark sector 
includes particles whose lifetimes can in principle collectively span 
a vast range of timescales from well before to long after the present day. 
Understanding the impacts of such decays for present-day cosmic-ray physics is 
therefore of paramount importance.

Because our goal in this paper is to explore the cosmic-ray phenomenology to which
DDM ensembles can give rise, we shall avoid focusing on a specific DDM model and instead
assume a general ensemble configuration of individual dark-matter components $\phi_n$
whose masses $m_n$ are given by a relation of the form
\begin{equation}
  m_n ~=~ m_0 + n^\delta \Delta m~,
\label{eq:MassSpectrum}
\end{equation} 
where the mass splitting $\Delta m$ and scaling exponent $\delta$ are both assumed positive.
Thus the index $n=0,1,2,...$ labels the ensemble constituents in order of increasing mass.  
Likewise, we shall assume that these components $\phi_n$ have
cosmological abundances $\Omega_n$ and decay widths $\Gamma_n$ which can be parametrized according
to general scaling relations of the form  
\begin{equation}
  \Omega_n ~=~ \Omega_0 \left(\frac{m_n}{m_0}\right)^{\alpha} ~,~~~~
  \Gamma_n ~=~ \Gamma_0 \left(\frac{m_n}{m_0}\right)^{\gamma}~
\label{eq:GammaOmegaScaling}
\end{equation}
where $\alpha$ and $\gamma$ are general power-law exponents.
While the existence of such scaling relations is not a fundamental requirement 
of the DDM framework, relations such as these do arise 
naturally in a number of explicit realistic DDM models~\cite{DynamicalDM1,DynamicalDM2,DDMColliders}
and allow us to encapsulate the structure of an entire DDM ensemble in terms of
only a few well-motivated parameters.
Note that the decay width $\Gamma_n$ in Eq.~(\ref{eq:GammaOmegaScaling}) refers to 
(or is otherwise assumed to be dominated by)
the decay of $\phi_n$ to SM states, and likewise 
$\Omega_n$ denotes the cosmological abundance 
that $\phi_n$ {\it would have had}\/ at the present time if it had been absolutely stable. 
Indeed, because the DDM framework allows each individual $\phi_n$ component to decay at a different time,
the corresponding abundances $\Omega_n$ generally evolve in a non-trivial manner 
across the DDM ensemble~\cite{DynamicalDM1}, and thus no single scaling relation can hold across 
the ensemble for all times.  

Given these scaling relations, our DDM ensemble is in principle described by
the seven parameters 
$\lbrace \alpha,\gamma,\delta,m_0,\Omega_0,\Gamma_0,\Delta m \rbrace$.
For convenience, in this paper we shall fix $\delta=1$ and $\Delta m=1$~GeV;
these choices ensure that our DDM ensemble transcends a mere set of 
individual dark-matter components and observationally acts as a ``continuum'' of states
relative to the scale set by the energy resolution of the relevant cosmic-ray detectors.   
We shall also fix $\Omega_0$ by requiring that the ensemble carry the entire observed 
dark-matter abundance $\OmegaDM$; this will be discussed further below.  Of the 
remaining four parameters, we shall treat $\lbrace \alpha,\gamma,m_0\rbrace$ as free
parameters and eventually survey over different possibilities 
within the ranges $-3 \leq \alpha < 0$, $-1\leq \gamma \leq 2.5$, and 
$100~{\rm GeV}\lsim m_0\lsim 1~{\rm TeV}$.  Given a specific assumption for how 
each $\phi_n$ decays to Standard-Model states,
we will then find that each such choice of $\lbrace \alpha,\gamma,m_0\rbrace$ leads
to a unique prediction for the overall shape of the resulting electron and positron
fluxes as functions of energy, with an arbitrary normalization set by the lifetime
$\tau_0\equiv 1/\Gamma_0$ of the lightest dark-matter component in the ensemble.
For each choice of $\lbrace \alpha,\gamma,m_0\rbrace$,
the final remaining parameter $\tau_0$ can therefore be determined through a best-fit analysis,
and indeed we shall find that most scenarios of interest have $\tau_0\gsim 10^{26}$~s.
Thus, in this paper, the three quantities $\lbrace \alpha,\gamma,m_0\rbrace$ shall
serve as our independent degrees of freedom parametrizing our DDM ensemble.
 
There are also additional phenomenological considerations which can be used to place 
bounds on these parameters.  For example, one generic feature of the DDM framework is 
an expected balancing of decay widths against abundances across the DDM ensemble.  
This expectation comes from the general observation that the earlier a dark-matter 
component might decay during the evolution of the universe, the smaller its cosmological 
abundance must be in order to avoid the disruptive effects of that decay and remain 
phenomenologically viable~\cite{DynamicalDM1,DynamicalDM2,DynamicalDM3}.  We therefore 
expect to find, roughly speaking, an {\it inverse}\/ relation between cosmological
abundances and decay widths, or equivalently that $\alpha \gamma <0$.  Indeed, as 
indicated above, it is usually $\alpha$ which will be negative in most 
DDM scenarios, while $\gamma$ is generally positive.  However, for illustrative 
purposes, in this paper we shall also occasionally consider extrapolations into 
regions of parameter space with $\alpha\gamma >0$.

Likewise, in this paper we also shall focus on regions of parameter space in which
$\gamma$ is not too large.  Our reasons, again, are primarily phenomenological.
In general, our interest in this paper concerns the contributions that the dark-matter 
components $\phi_n$ might, through their decays, make to the differential 
electron/positron fluxes $\Phi_{e^\pm}$ within the energy range 
$20~{\rm GeV} \lesssim E_{e^\pm} \lesssim 1$~TeV.~  One of the most interesting regions 
of parameter space will therefore be that in which all of the $\phi_n$ which could in 
principle yield a non-negligible contribution to these fluxes are sufficiently 
long-lived that their abundances $\Omega_n$ are effectively undiminished by decays 
and consequently still scale according to Eq.~(\ref{eq:GammaOmegaScaling}) at the 
present time.  Indeed, this is the regime within which the full DDM ensemble plays
the most significant role in indirect-detection phenomenology and within which the 
most distinctive signatures arise.  In order to specify where this regime lies within 
the parameter space of our DDM model, we begin by noting that the contribution from 
extremely heavy dark-matter components $\phi_n$ to $\Phi_{e^\pm}$ will be 
comparatively negligible (\ie, below background) for 
$E_{e^\pm} \lesssim 1$~TeV.  ~We therefore define a fiducial mass scale $m_\ast$
to represent this cutoff, and demand that all components $\phi_n$ with masses 
$m_n < m_\ast$ have lifetimes $\tau_n > \tnow$, where $\tnow \approx 4.3 \times 10^{17}$~s 
is the age of the universe.  The scaling relation for $\Gamma_n$ in 
Eq.~(\ref{eq:GammaOmegaScaling}) implies that this condition may be written as a 
constraint on the scaling exponent $\gamma$:
\begin{equation}
  \gamma ~\lesssim~ \frac{\ln \big(\tau_0/\tnow\big)}{\ln\big(m_\ast/m_0\big)}~.
\label{eq:GammaConsistencyCondit}
\end{equation}
For any ensemble with $\tau_0 \gtrsim 10^{26}$~s and $m_0\gtrsim 200$~GeV,
we find that the conservative choice $m_\ast = 10^6$~GeV yields the limit
$\gamma \lesssim 2.26$.
As we shall see in Sect.~\ref{sec:results}, it is not difficult to satisfy
the condition in Eq.~(\ref{eq:GammaConsistencyCondit}) while simultaneously
reproducing the positron-fraction curve reported by \mbox{AMS-02} and satisfying
all other applicable constraints.
However, we hasten to emphasize that the criterion in Eq.~(\ref{eq:GammaConsistencyCondit})
does {\it not}\/ represent a parameter-space constraint which our DDM model must satisfy
for theoretical or phenomenological consistency.
By contrast, it merely defines a regime of particular phenomenological interest
within our model.

Finally,
in order to evaluate the contribution to the observed cosmic-ray electron and positron 
fluxes from such a DDM ensemble, it is necessary to specify not only the 
particle-physics properties of the DDM 
ensemble itself but also certain astrophysical properties of the ensemble 
which characterize how the populations of the various $\phi_n$ are distributed 
throughout the galactic halo.  First, for this analysis, we assume that the DDM ensemble 
contributes essentially the entirety of the present-day dark-matter abundance, and 
therefore that the total DDM ensemble abundance 
\begin{equation}
  \Omegatot ~\equiv~ \Omega_0 \sum_{n=0}^{n_{\mathrm{max}}} 
  \left(1 + n^\delta \frac{\Delta m}{m_0}\right)^\alpha
\label{eq:Omegatot}
\end{equation}
matches the total dark-matter abundance 
$\OmegaDM h^2 \approx  0.1131 \pm 0.0034$ observed by 
WMAP~\cite{WMAP}.
  (Although recent Planck results~\cite{PlanckResults} suggest the slightly 
  higher value $\OmegaDM h^2 \approx 0.1199 \pm 0.0027$,
  such a shift in $\OmegaDM$ has an essentially negligible effect on our results.)
Note that since we are considering
only those DDM ensembles for which 
$\alpha\delta < -1$, the sum in Eq.~(\ref{eq:Omegatot}) 
remains convergent in the $n_{\mathrm{max}}\rightarrow \infty$ limit~\cite{DDMDirectDet}. 
Second, we make the simplifying
assumption that the density profiles $\rho_n(r)$ according to which our
individual dark-matter components $\phi_n$ are distributed within the galactic halo share
a common functional form, which we model using an NFW profile~\cite{NFW}.  
Finally, for simplicity, 
we take the normalization of each $\rho_n(r)$ within the 
galactic halo --- which is typically specified by the local dark-matter density 
$\rho^{\mathrm{loc}}_n$ within the solar neighborhood --- to be proportional to the global 
energy density of the corresponding constituent.  Thus, we shall assume
that $\rho^{\mathrm{loc}}_n/\rho_{\mathrm{tot}}^{\mathrm{loc}} = \Omega_n/\Omegatot$
in what follows, with $\rhototloc \approx 0.3\mathrm{~GeV}/\mathrm{cm}^3$.


\section{Electron/Positron Production and Propagation\label{sec:propagation}}

  
In general, a given dark-matter particle $\phi$ undergoes a decay of the form 
$\phi\to f$ where $f$ is a multi-particle final state which includes at least 
some visible-sector fields.  Using only visible-sector physics, one can then 
extract a set of differential electron and positron fluxes 
$dN_{f,e^\pm}/d E_{e^\pm}$ which reflect not only the kinematics of how $f$ 
subsequently decays to electrons and positrons, but also the possible decay 
chains and branching ratios that might be involved in such subsequent processes, 
the effects of final-state radiation, and so forth.  However, 
$dN_{f,e^\pm}/d E_{e^\pm}$ only describes the differential fluxes at the location 
where these electrons and positrons were originally produced (here assumed to 
be somewhere within our galaxy); it is still necessary to use this so-called 
``injection'' spectrum $dN_{f,e^\pm}/d E_{e^\pm}$ in order to determine the final 
electron and positron differential fluxes $\Phi_{e^\pm}$ that will emerge and be 
measured after these particles have propagated through the interstellar 
medium (ISM) and entered our solar neighborhood.  In this section, we shall 
discuss how these final observed differential fluxes $\Phi_{e^\pm}$ may be 
determined, focusing on the case when our injection spectrum 
$dN_{f,e^\pm}/d E_{e^\pm}$ arises from an entire DDM ensemble.  Note, in 
particular, that our interest in this paper centers on the 
{\it differential}\/ fluxes $\Phi_{e^\pm}$;   these quantities are
related to the {\it total}\/ fluxes $\tilde \Phi_{e^\pm}$ via 
$\Phi_{e^\pm}\equiv d\tilde \Phi_{e^\pm}/d E_{e^\pm}$.  Indeed, it is only the 
differential fluxes $\Phi_{e^\pm}$ which exhibit the all-important 
energy-dependence which is ultimately the focus of our analysis.  In this 
vein, we also note that we are only concerned in this paper with total 
differential fluxes integrated over all angles, and not with their
directional dependence.

In general, these differential fluxes $\Phi_{e^\pm}$ can be expressed as 
\begin{equation}
  \Phi_{e^\pm} ~=~ \frac{v}{4\pi} f_{e^\pm}(E)~
\label{eq:RelFluxAndPhaseSpaceDensity}
\end{equation}
where $f_{e^\pm}(E)$ denotes the local differential number density of electrons and positrons
per unit energy and where $v \approx c$ denotes the velocity of the incident 
particles.  Taken as a function of position, energy, and time,
this number density $f_{\pm}(E,\vec{r},t)$ is in turn determined by the transport equation
\begin{eqnarray}
   \frac{\partial f_{e^\pm}}{\partial t} &=&
  \vec{\nabla} \cdot \left[K(E,\vec{r})\vec{\nabla} f_{e^\pm}\right] \nonumber\\ 
  && ~~~~ + \frac{\partial}{\partial E} \Big[b(E,\vec{r})f_{e^\pm}\Big] + 
  Q_{e^\pm}(E,\vec{r},t)~,~~~~
\label{eq:TransportEquation}
\end{eqnarray}
where $Q_{e^\pm}(E,\vec{r},t)$ is the source term for electron and positron 
production, where $K(E,\vec{r})$ is the diffusion coefficient, and where $b(E,\vec{r})$ 
is the energy-loss rate.  For an approximately steady-state process,
we may take $Q_{e^\pm}(E,\vec{r},t) \approx Q_{e^\pm}(E,\vec{r})$ as effectively 
independent of time, and thus we have $\partial f_{e^\pm}/\partial t = 0$.
Of course, the total injection rate $d N_{f,e^\pm}/dE_{e^\pm}$ associated with
the decaying constituents of a DDM ensemble will by nature be time-dependent.  
However, for ensembles capable of producing a non-negligible contribution to 
observed electron and positron fluxes, we shall see that the timescale on which 
this variation is significant is far greater than the timescale for $e^{\pm}$ 
diffusion through the galactic halo.  Thus this steady-state approximation is 
justified.

Following Ref.~\cite{IbarraTran1}, we next adopt a stationary 
two-zone diffusion model in which the diffusion coefficient is spatially constant 
throughout the diffusion zone and takes the form
\begin{equation}
  K(E) ~=~ {v\over c}\, K_0\, \mathcal{R}^{\epsilon}~, 
  \label{eq:DiffusionCoefficient}
\end{equation} 
where $K_0$ and $\epsilon$ are free parameters which characterize a particular 
diffusion model and where
$\mathcal{R}$ is the so-called ``rigidity'' of the particle 
(defined as the ratio of its momentum in GeV to its electromagnetic charge in units
of the electron charge $e$).  Note that for electrons and positrons with 
$E \gg m_e \approx 511$~keV, the diffusion coefficient may also be expressed as
$K(E) \approx K_0 (E/\mathrm{GeV})^\epsilon$.  
The energy-loss rate $b(E,\vec{r})$ includes contributions from both synchrotron 
radiation and inverse-Compton scattering and can be written in the form
\begin{equation}
  b(E,\vec{r}) ~=~ \frac{32\pi \alpha_{\mathrm{EM}}^2}{9m_e^4} E^2
    \left[u_B(\vec{r}) + \sum_i u_{\gamma,i}(\vec{r})R_i^{\mathrm{KN}}(E)\right]~,
\end{equation} 
where $u_B(\vec{r})$ is the energy density in galactic magnetic fields; where the 
$u_{\gamma,i}(\vec{r})$ are the contributions to the photon energy density 
from the CMB, starlight, and diffuse infrared light ($i=1,2,3$, respectively); 
and where the functions $R^{\mathrm{KN}}_i(E)$ describe 
the energy dependence of the corresponding contributions from these three sources.
The functional forms for $u_B(\vec{r})$, the $u_{\gamma,i}(\vec{r})$, and the
$R^{\mathrm{KN}}_i(E)$ can be found in Ref.~\cite{PPPC4DMID} and references therein.
Finally, the diffusion zone is assumed to be cylindrical, with a radius $R_D$ and 
half height $L_D$.  
For this analysis we adopt the so-called ``MED'' propagation model of 
Refs.~\cite{DelahayePositronPropagation,DonatoAntiprotonPropagation} in which 
$\epsilon = 0.70$, $\mathcal{K}_0 = 0.0112~{\rm kpc}^2$/Myr, $L_D = 4$~kpc, and $R_D = 20$~kpc.  
Other choices will be discussed in Sect.~\ref{sec:varyinginputs}.
  
In general, it is the source terms 
$Q_{e^\pm}(E,\vec{r})$ which encode the specific dark-matter model under scrutiny
and its possible decay patterns.
For a DDM model consisting of an ensemble of dark-matter components $\phi_n$, 
the source terms $Q_{e^\pm}(E,\vec{r})$ for electrons and
positrons take the general form
\begin{equation}
  Q_{e^\pm}(E,\vec{r}) ~=~ \sum_{n=0}^{n_{\max}}\frac{\rho_n(\vec{r})}{m_n}
  \Gamma_n \sum_f \mathrm{BR}(\phi_n \rightarrow f) \frac{dN^{(n)}_{f,e^\pm}}{dE_{e^\pm}}~,
  \label{eq:DDMepemSourceTerm}
\end{equation}  
where $\rho_n$ denotes the energy density of the DDM component $\phi_n$, 
where $\mathrm{BR}(\phi_n \rightarrow f)$ denotes the branching fraction for the decay
$\phi_n\to f$, and where $dN^{(n)}_{f,e^\pm}/dE'_{e^\pm}$ are the differential injection
spectra produced by each such decay.  
The solution to Eq.~(\ref{eq:TransportEquation}) 
can therefore be expressed in the form 
\begin{widetext}
\begin{equation}
   \Phi_{e^\pm}^{\mathrm{DDM}}(E_\pm, \vec r) ~\approx~ 
  \frac{c}{4\pi} \sum_{n=0}^{n_{\mathrm{max}}} 
    \frac{\Gamma_n}{m_n} 
  \int d^3 \vec r\,' \, \rho_n(\vec r\,')
    \int_0^{m_n/2} dE_{e^\pm}'  
   \,  G_{e^\pm}(E_{e^\pm},E_{e^\pm}';\vec r,\vec r\,')\,
    \sum_f \mathrm{BR}(\phi_n \rightarrow f) \frac{dN^{(n)}_{f,e^\pm}}{dE_{e^\pm}'}(E_{e^\pm}')~,~
\label{eq:FluxIntegralDDMGen}
\end{equation}  
where $G_{e^\pm}(E_{e^\pm},E_{e^\pm}';\vec r,\vec r\,')$ is the Green's function solution to 
Eq.~(\ref{eq:TransportEquation}).  
Indeed, this equation indicates that the differential flux which 
results from the decaying DDM ensemble is nothing but the sum of the individual
differential fluxes which would have resulted from the decays of each DDM component individually ---
precisely as expected for an essentially linear propagation model wherein
the Green's function $G_{e^\pm}(E_{e^\pm},E_{e^\pm}'; \vec r, \vec r\,')$ 
encapsulates the essence of propagation through the interstellar medium.

For a DDM ensemble parametrized as in
Sect.~\ref{sec:ensembles} --- and under the assumption that the galactic energy densities 
$\rho_n(\vec r\,')$ are approximately proportional to the
corresponding global energy densities $\rho_n = \Omega_n \rhocrit$ --- 
the expression in Eq.~(\ref{eq:FluxIntegralDDMGen}) takes the form
\begin{eqnarray}
  \Phi_{e^\pm}^{\mathrm{DDM}} &\approx& \frac{c\, 
      \Omega_0}{4\pi\,\Omegatot \,\tau_0 \,m_0}
   \sum_{n=0}^{n_{\mathrm{max}}} 
    \left(1 + n^\delta\frac{\Delta m}{m_0}\right)^{\alpha + \gamma - 1}
   \int d^3 \vec r\,'   \,\rho_{\rm tot}(\vec r\,') \nonumber\\
   && ~~~~~~~~~~~\times~ 
    \int_0^{(m_0+ n^\delta \Delta m)/2} dE_{e^\pm}'\, 
    G_{e^\pm}(E_{e^\pm},E_{e^\pm}';\vec r,\vec r\,') \, \sum_f \, \mathrm{BR}(\phi_n \rightarrow f)
    \, \frac{dN^{(n)}_{f,e^\pm}}{dE_{e^\pm}'}(E_{e^\pm}')~,
\label{eq:FluxIntegralDDMParam}
\end{eqnarray}  
\end{widetext}
where $\Omegatot$ is defined in Eq.~(\ref{eq:Omegatot}) and
where $\tau_0$ once again denotes the lifetime of the lightest ensemble constituent.
In practice, we model the energy densities $\rho_n(\vec r)$ [and thus $\rho_{\rm tot}(\vec r)$] to
be spatially distributed according to the NFW halo profile~\cite{NFW},
and we evaluate the expressions in Eq.~(\ref{eq:FluxIntegralDDMParam}) 
numerically, using the publicly available PPPC4DMID package~\cite{PPPC4DMID} to 
determine the electron and positron spectra at injection as well as to 
determine the effects of propagating these injected particles 
through the interstellar medium.  However, as a cross-check, we have also 
verified that the resulting differential fluxes agree with the analytic results 
obtained using the approximate analytic Green's function~\cite{IbarraTran1} 
corresponding to the same choice of propagation model.

It is important to note the manner in which the DDM model parameters 
$\alpha$, $\gamma$, and $\tau_0$ appear in the expression in 
Eq.~(\ref{eq:FluxIntegralDDMParam}).  In particular, it is only the 
combination $\alpha + \gamma$ which appears in the summand, as this combination 
dictates how the injected flux of $e^\pm$ due to $\phi_n$ decays scales across 
the ensemble.  In so doing, this combination determines the {\it shape}\/ of 
the observed $e^\pm$ flux spectra.
Of course, the exponent $\alpha$ also implicitly appears in the overall normalization 
prefactor, since it affects the value of $\Omega_0$ when $\Omegatot$ is set
equal to $\OmegaDM$.  However, any change in this normalization factor due to a 
change in $\alpha$ can be absorbed into a corresponding rescaling  
of $\tau_0$.  Thus only $\alpha+\gamma$ and $\tau_0$ serve as independent degrees of freedom
insofar as the electron/positron fluxes are concerned.
We will therefore express our results in terms of the 
combination $\alpha + \gamma$ in what follows, and perform a best-fit analysis to \mbox{AMS-02}  
data in order to fix $\tau_0$ for any choice of $\alpha+\gamma$.
The particulars of this analysis will be discussed in Sect.~\ref{sec:results}.

In addition to the primary contributions to $\Phi_{e^\pm}$ from dark-matter
decay, we must also take into account the background contribution to these fluxes from 
astrophysical processes.  
In principle, these fluxes are
specified by the choice of propagation model and the injection spectrum of
$e^\pm$ from astrophysical sources, including both a primary contribution
from objects such as supernova remnants and a secondary component due to
the spallation of cosmic rays on the interstellar
medium. In practice, however, the injection spectrum is not well known, and
thus specifying a propagation model is still not sufficient to determine
the astrophysical background fluxes of electrons and positrons at the
location of the Earth.  For this reason, following, {\it e.g.}\/, 
Refs.~\cite{IbarraThreeBody,IbarraTranWeniger,CosmicThreeBodyRayDM}, we
adopt a background-flux model which provides a reasonably good empirical
fit to the observed fluxes at low $E_{e^\pm}$, namely 
the so-called ``Model~0'' presented by the FERMI collaboration in
Ref.~\cite{FERMIPosDataBGModels}.  
These background fluxes are well described 
by the parametrizations~\cite{IbarraTranWeniger}
\begin{widetext}
\begin{eqnarray}
  \Phi^{\mathrm{BG}}_{e^-} & \approx & k \left(10^{-4}\right)\times 
     \Bigg[\frac{82.0\, (E_{e^-}/\mathrm{GeV})^{-0.28}}
     {1+0.224\, (E_{e^-}/\mathrm{GeV})^{2.93}}\Bigg] 
     \mathrm{~GeV}^{-1}\mathrm{\,cm}^{-2}\mathrm{\,s}^{-1}\mathrm{\,sr}^{-1}
   \nonumber \\ \rule[-1em]{0pt}{1.2cm}
  \Phi^{\mathrm{BG}}_{e^+} & \approx & \phantom{k} \left(10^{-4}\right) \times 
     \Bigg[\frac{38.4\, (E_{e^+}/\mathrm{GeV})^{-4.78}}
     {1+0.0002\, (E_{e^+}/\mathrm{GeV})^{5.63}} 
     + 24.0\, (E_{e^+}/\mathrm{GeV})^{-3.41}\Bigg]
   \mathrm{~GeV}^{-1}\mathrm{\,cm}^{-2}\mathrm{\,s}^{-1}\mathrm{\,sr}^{-1}~,
  \label{eq:BGFluxesFunctionalForms}
\end{eqnarray}
\end{widetext}
where $k$ is a normalization coefficient which parametrizes the uncertainty
in the background $e^-$ flux. In our analysis, we allow $k$ to fluctuate
within the range $0.7\leq k\leq≤1.0$.  This single degree of freedom clearly does
not parametrize all of the uncertainties in the background fluxes.  However, it does
provide some measure of flexibility for these fluxes which will be sufficient
for our purposes.  Indeed, although ``Model 0'' 
(which we use for calculating the astrophysical backgrounds) is quite different from 
the MED model (which we use to calculate those fluxes which originate from our
DDM ensemble), ``Model 0'' has the benefit that it successfully fits the measured
background $e^\pm$ flux spectra in a suitable low-energy ``control'' region where data
actually exists.   
A more complete discussion of the effects of uncertainties in
the astrophysical background flux can be found in Ref.~\cite{GaggeroMaccione}.
We will also discuss the treatment of these fluxes further in Sect.~\ref{sec:varyinginputs}.

Finally, we remark that these interstellar background fluxes can be 
significantly modified by solar modulation effects at very
low energies $E_{e^\pm} \lesssim 10$~GeV.~  
Indeed, for $E_{e^\pm} \lesssim 10$~GeV,  the observed flux spectra at the top
of the atmosphere can differ considerably from the functional forms given in 
Eq.~(\ref{eq:BGFluxesFunctionalForms}).  However,
our main interest in this paper concerns the significantly higher energy range
$20~{\rm GeV}\lsim E_{e^\pm}\lsim 1~{\rm TeV}$.
Therefore, we shall disregard the effects of solar modulation in most of
what follows.  However, in all figures displayed in this paper, the results 
shown actually include this modulation effect, which we have calculated using the
so-called force-field approximation~\cite{ForceField}.  Under this approximation,
the observed fluxes are related to the interstellar fluxes via the modification~\cite{Perko}
\begin{equation}
  \Phi_{e^\pm}^{\mathrm{BG,obs}} ~=~
    \left(\frac{E_{e^\pm}}{E_{e^\pm}+e\phi_F}\right)^2
    \Phi_{e^\pm}^{\mathrm{BG}}(E_{e^\pm} + e\phi_F)~,
  \label{eq:BGElecPosFluxesSolarMod}
\end{equation}
where $e$ is the electron charge and where $\phi_F = 550$~MeV is the value we adopt
for the solar-modulation potential. 
A more complete discussion of solar-propagation 
modelling can be found in Ref.~\cite{solarmodulation}.

In summary, the total differential fluxes of cosmic-ray electrons and positrons in our
DDM model is given by the sum of the corresponding signal contribution in 
Eq.~(\ref{eq:FluxIntegralDDMParam}) and the background contribution in 
Eq.~(\ref{eq:BGElecPosFluxesSolarMod}):    
\begin{equation}
  \Phi_{e^\pm} ~=~ \Phi_{e^\pm}^{\mathrm{DDM}} + \Phi_{e^\pm}^{\mathrm{BG,obs}}~.  
  \label{eq:TotalElecPosFluxes}
\end{equation}
Given these fluxes,
the combined flux $\Phi_{e^+} + \Phi_{e^-}$ and the positron fraction
$\Phi_{e^+}/(\Phi_{e^+} + \Phi_{e^-})$ directly follow.


\section{Phenomenological Constraints\label{sec:constraints}}


In this section, we discuss the phenomenological constraints that we shall
impose on our DDM model.
In particular, we shall require that our general DDM model be consistent with
\begin{itemize}
\item  limits from the PAMELA experiment~\cite{PAMELAAntiproton} 
        on the flux of cosmic-ray antiprotons;
\item  limits from the FERMI-LAT experiment~\cite{FERMIGammaRays} on the 
        observed gamma-ray flux, and especially on its diffuse 
        isotropic component;
\item  constraints~\cite{SynchrotronDecayDMLimits} 
         on the synchrotron radiation produced via the interaction between
          high-energy electrons and positrons within the galactic halo and 
          background magnetic fields;
\item  constraints on the ionization history of the universe, as 
         recorded in the CMB, from existing anisotropy data and anticipated
         polarization data from Planck~\cite{PlanckResults}; and
\item  constraints from the FERMI-LAT experiment~\cite{FERMIPositronData2010}
           on the combined $e^\pm$ flux.
\end{itemize}
We shall now discuss each of these in turn.

\subsection{Cosmic-ray antiproton constraints}

As discussed in the Introduction, limits from
PAMELA~\cite{PAMELAAntiproton} on the flux of cosmic-ray antiprotons
impose non-trivial constraints on dark-matter models which purport to 
explain the positron excess.  These constraints are particularly stringent 
for dark-matter candidates which decay primarily either directly into 
quarks or gluons, or else into $W^\pm$ or $Z$ bosons which in turn produce 
such particles with significant branching fractions via their subsequent decays.

However, these constraints are far less stringent for dark-matter candidates which 
decay primarily into charged leptons.  For this reason, in our analysis we 
shall focus primarily on the case in which the constituents of the DDM ensemble 
decay leptonically, via processes of the form $\phi_n \rightarrow \ell^+\ell^-$, 
where $\ell = \{e,\mu,\tau\}$.
In each case, we have 
used the PPPC4DMID package~\cite{PPPC4DMID}  
to calculate the contribution to the cosmic-ray antiproton flux
from the leptonic decays of our DDM ensemble,
and we have verified that
the antiproton flux lies well below experimental limits for all 
relevant antiproton energies.  
Indeed, this conclusion holds
within all portions of the DDM parameter space which 
ultimately prove relevant for explaining the positron excess.

We emphasize that while other similar limits on decaying dark-matter 
particles exist --- for 
example, constraints on the contributions of such particles to cosmic-ray 
antideuteron fluxes --- these constraints do not significantly impact the parameter 
space of DDM ensembles whose constituents decay primarily via leptonic channels.

\subsection{Gamma-ray flux constraints}
    
The flux of gamma rays produced from annihilating or decaying dark-matter particles 
within the galactic halo is also tightly constrained by observation, as is the contribution 
from a cosmological population of decaying dark-matter particles to the
isotropic gamma-ray flux.  The latter constraints are typically more 
stringent for decaying dark-matter models~\cite{CirelliIsoAndOtherGammaRay,CirelliDiffGammaRayBounds}; 
moreover, they do not depend on the
halo profile or other unknown properties of the dark-matter distribution within our
galaxy.  We therefore focus here on the isotropic gamma-ray constraints.

In general, the total contribution to the apparent isotropic
gamma-ray flux from dark-matter decay 
receives two sub-contributions:
\begin{equation}
  \Phi_\gamma^{\mathrm{Iso}} ~=~ \Phi_\gamma^{\mathrm{EGB}} + 
  4\pi\left.\frac{d\Phi_\gamma^{\mathrm{DGB}}}{d\Omega}\right|_{\mathrm{min}}~.
  \label{eq:DecomposeIsoFlux}
\end{equation}
The first of these is the contribution from a cosmological population of decaying 
dark-matter particles to the true diffuse extragalactic gamma-ray flux.  The second 
is the isotropic component of the residual contribution from decaying dark-matter 
particles in the galactic halo.  This latter contribution includes 
individual contributions from prompt gamma-ray production and from gamma-ray 
production via the inverse-Compton scattering of $e^\pm$ produced by $\phi_n$ decays 
off background photons.  We evaluate both of these contributions to the residual
galactic background flux, as well as the truly diffuse extragalactic flux contribution,
using the PPPC4DMID package~\cite{PPPC4DMID}.  Following the analysis in 
Ref.~\cite{CirelliDiffGammaRayBounds}, 
we assume that the direction at which this latter contribution reaches 
a minimum is that
opposite the galactic center.  

Since there is substantial uncertainty in the background contribution to the 
isotropic gamma-ray flux from astrophysical sources (see, \eg, the various 
possible contributions discussed in Ref.~\cite{GammaFluxUncert}),    
we require as a consistency condition only that the contribution to this flux 
predicted from the decay of a given DDM ensemble alone not exceed the flux 
reported by the FERMI collaboration~\cite{FERMIGammaRays}.  In this manner, 
given a particular choice of model parameters, we determine a lower bound on 
the lifetime $\tau_0$ of the lightest ensemble constituent.
We do this by computing the goodness-of-fit statistic 
\begin{equation}
  \chi^2 ~=~ \sum_{i=1}^N \frac{\big(\Phi_i^{\mathrm{obs}} - 
  \Phi_i^{\mathrm{DDM}}\big)^2}
  {\big(\Delta \Phi_i^{\mathrm{obs}}\big)^2}\,
  \Theta\big(\Phi_i^{\mathrm{DDM}} - \Phi_i^{\mathrm{obs}}\big)~,
  \label{eq:ChiSqForFERMI}
\end{equation}      
where the index $i$ labels the energy bins into which the FERMI data are partitioned, 
where $\Phi_i^{\mathrm{DDM}}$ is the differential gamma-ray flux for bin $i$ predicted by the 
DDM model in question, where $\Phi_i^{\mathrm{obs}}$ is the central value reported 
by FERMI for the gamma-ray flux in the corresponding bin,  
where $\Delta \Phi_i^{\mathrm{obs}}$ is the uncertainty in that central 
value, and where $\Theta(x)$ denotes the Heaviside theta function.  
We then compute a (one-sided) $p$-value by comparing 
this goodness-of-fit statistic to a $\chi^2$ 
distribution with $N$ degrees of freedom, where $N = 9$ is the number of energy bins 
used in the FERMI gamma-ray analysis.  Finally, we express this result in terms of the number of 
standard deviations away from the mean to which this $p$-value would correspond for 
a (two-sided) Gaussian distribution.  In this analysis, we adopt as our criterion for
consistency with FERMI data the requirement that the isotropic gamma-ray flux contributed
by the decays of the DDM ensemble constituents agree with the FERMI data 
to within $3\sigma$.    
We note, however, that there also exist other methods~\cite{otherFermi} of using FERMI gamma-ray data 
to constrain the properties of decaying or annihilating dark matter.

\subsection{Synchrotron radiation constraints}

Gamma-ray signatures of this sort are not the only way in which
photon signals constrain the properties of the dark matter.
Indeed, synchrotron radiation produced via the interaction between
high-energy electrons and positrons within the galactic halo
and background magnetic fields can result in an observable
radio signal.  Observational limits on such a signal therefore
constrain scenarios in which electrons and positrons are produced by
a population of annihilating~\cite{SynchrotronGeneralAnn} or
decaying~\cite{SynchrotronGeneralDec,SynchrotronDecayDMLimits}
dark-matter particles.  Constraints of this sort were derived in
Ref.~\cite{SynchrotronDecayDMLimits} for the case of a traditional
dark-matter candidate $\chi$ which decays primarily into charged
leptons, and in particular into $e^+e^-$ pairs.  For the choice of
halo profile and propagation model we adopt here, it was shown that
these constraints are generically subleading in comparison with
direct constraints on the positron fraction for $m_\chi \gtrsim 11$~GeV.~
Indeed, the most stringent bound on the lifetime of any
leptonically-decaying $\chi$ with $m_\chi$ in this mass regime
from synchrotron-emission considerations alone is
$\tau_\chi \gtrsim 10^{26}$~s, again for a particle which decays
essentially exclusively to $e^+e^-$.  (The corresponding constraints
on a particle with a significant branching fraction to $\mu^+\mu^-$
or $\tau^+\tau^+$ are even less stringent~\cite{SynchrotronGeneralDec}.)
Since the above results hold for a single dark-matter component $\chi$ 
with $m_\chi\gtrsim 11$~GeV, they will necessarily hold 
component-by-component across our entire DDM ensemble so long as $m_0\gtrsim 
11$~GeV and $\alpha + \gamma < 0$.  As we shall see, this latter condition 
is necessary in order 
to ensure that the injection energy decreases as a function of increasing 
mass within the ensemble.  We therefore conclude that 
the synchrotron-emission constraints will always be less 
stringent than the limits on $\Phi_{e^+}$ from FERMI and \mbox{AMS-02} data for 
any DDM ensemble satisfying these two constraints.

\subsection{CMB ionization history constraints}

In addition to direct limits on the observed gamma-ray and synchrotron fluxes,
dark-matter decays in the early universe are also constrained by considerations 
related to the CMB.~  In particular, high-energy photons, electrons, and positrons 
produced as a result of dark-matter decays in the early universe can alter the 
ionization history of the universe, thereby leaving an observable imprint on the 
CMB.~  That no such imprint has been observed in CMB data implies stringent 
constraints~\cite{ChenKamionkowski,PadmanabhanCMBPolLimits,
GalliCMBPolLimits,SlatyerCMBPolLimits,FinkbeinerSommerfeldEnhAndCMBPol,
SlatyerCMBConstraints} on dark-matter annihilation and decay.  Forthcoming
CMB-polarization data from Planck will improve upon these limits,
and projections based on the expected performance of the Planck satellite for the case
of a single long-lived dark-matter particle $\chi$ have been assessed by a number of 
authors~\cite{SlatyerCMBConstraints}.  

Broadly speaking, the results of 
these studies indicate that such a dark-matter particle $\chi$ must have a lifetime 
$\tau \gtrsim 10^{26}$~s if it is to have a cosmological abundance
$\Omega_\chi \sim \mathcal{O}(1)$, and that the upper limit on the 
present-day abundance of such a long-lived particle drops rapidly as
the lifetime of the particle decreases.  These studies also indicate that the upper
bound on $\Omega_\chi$ is not particularly sensitive to the mass of $\chi$.

\begin{figure*}[t]
\begin{center}
  \epsfxsize 2.75 truein \epsfbox {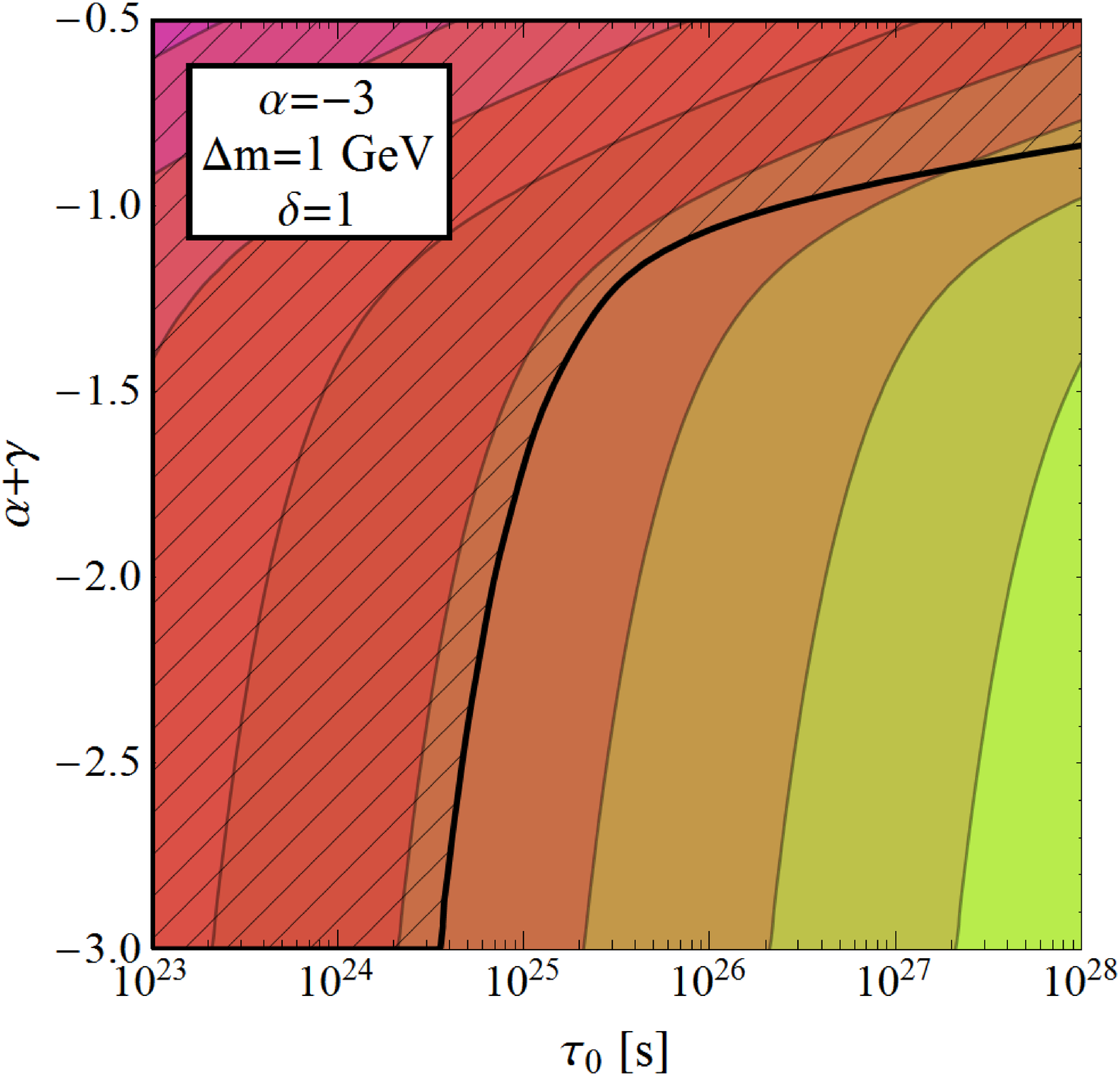} ~~
  \hskip 0.2 truein
  \epsfxsize 2.75 truein \epsfbox {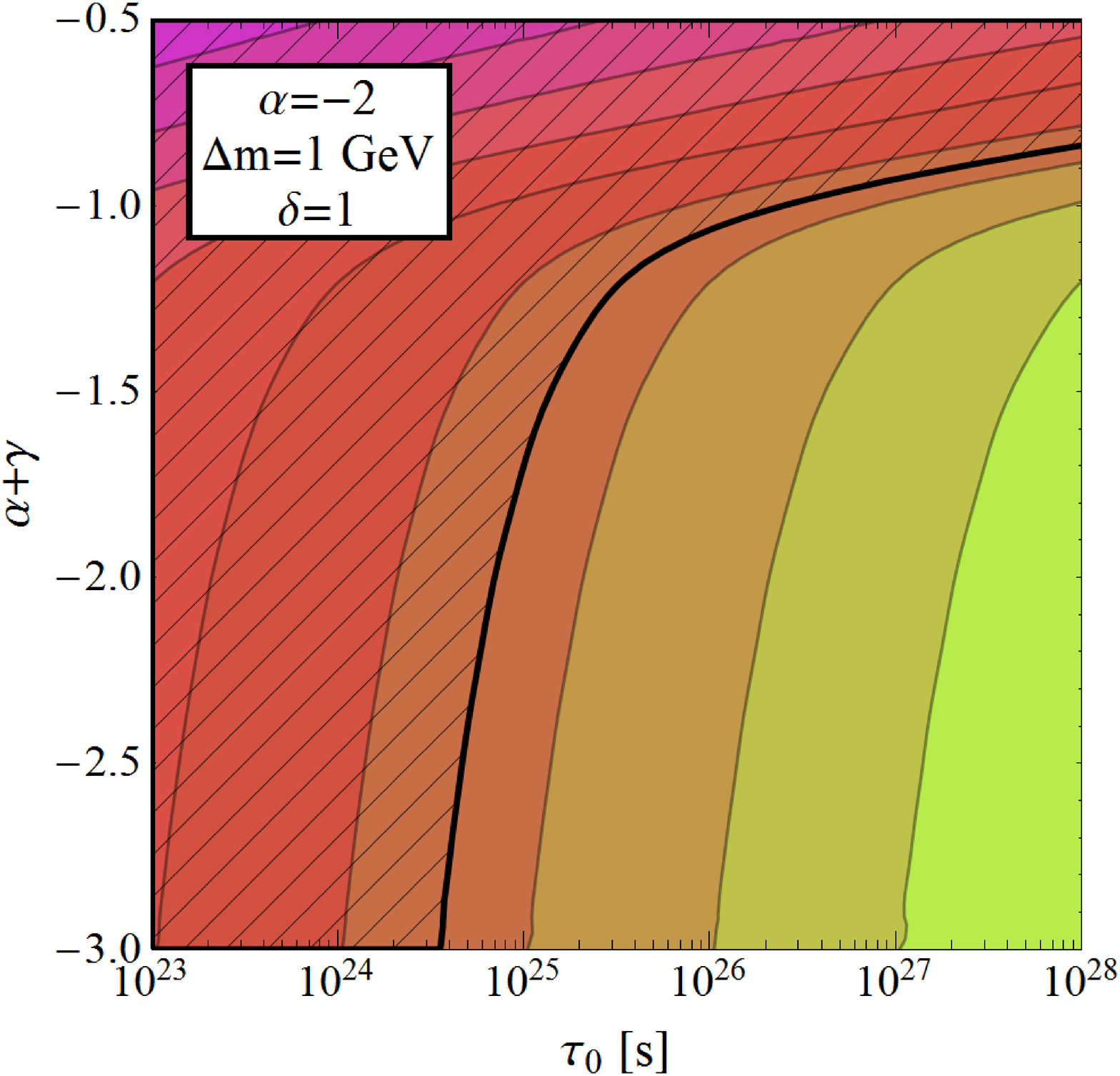}
  \\
  \raisebox{0.5cm}{\large $\xi \, [s^{-1}]$:~~}
    \epsfxsize 5.00 truein \epsfbox {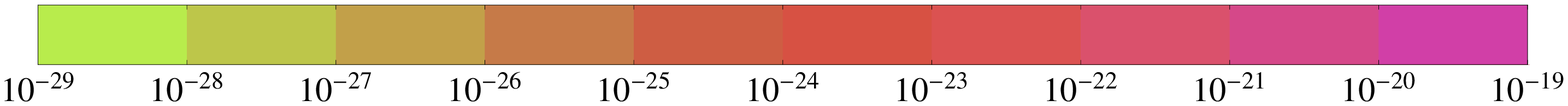}
\end{center}
\caption{The region of $(\tau_0,\alpha+\gamma)$ DDM parameter space excluded by 
  prospective CMB constraints on modifications of the ionization history of the 
  universe due to the presence of a decaying DDM ensemble 
  parametrized as in Eqs.~(\protect\ref{eq:MassSpectrum}) and 
  (\protect\ref{eq:GammaOmegaScaling}). 
  The results shown in the left and 
  right panels correspond respectively to $\alpha = -3$ and $\alpha = -2$.  
  The contours shown indicate the value of the injection-energy parameter 
  $\xi$, and the hatched region demarcated by the solid black curve is the 
  region of parameter space excluded by the CMB-consistency criterion 
  in Eq.~(\protect\ref{eq:CMBCriterionGeneralDDM}).     
\label{fig:CMBCriterion}} 
\end{figure*}

A precise determination of the corresponding limits on the parameter space 
of any given DDM scenario would require a detailed re-analysis of the ionization 
history of the universe in the presence of a DDM ensemble.  However, we can 
derive a rough criterion for consistency with CMB data from the usual CMB  
limits on decaying dark-matter particles based on the observation that such limits essentially 
constrain the energy injection from dark-matter 
decays~\cite{ChenKamionkowski}, and that these constraints are not particularly sensitive to the 
mass of the decaying particle.  Indeed, for any such particle with a lifetime $\tau_\chi \gtrsim 10^{13}$~s, 
the projected Planck limits derived in Ref.~\cite{SlatyerCMBConstraints} 
essentially amount to a constraint 
$\Omega_\chi \Gamma_\chi \lesssim 3 \times 10^{-26}\mathrm{~s}^{-1}$.
We can therefore establish a rough criterion for consistency with CMB data
by imposing an analogous condition on the total energy injection from the DDM 
ensemble as a whole: 
\begin{equation}
  \xi ~\equiv~ \sum_{n=0}^{n_{\mathrm{max}}}\Omega_n \Gamma_n 
  ~\lesssim ~3 \,\times \,10^{-26}\mathrm{~s}^{-1}~.
  \label{eq:CMBCriterionGeneralDDM}
\end{equation}
Indeed,
for the specific case of a DDM ensemble parametrized 
as in Sect.~\ref{sec:ensembles},
the quantity $\xi$ takes the form
\begin{eqnarray}
 && \xi ~=~ 
 \frac{\Omega_0}{\tau_0} \sum_{n=0}^{N_{\mathrm{CMB}}} 
 \left(1 + n^\delta\frac{\Delta m}{m_0}\right)^{\alpha + \gamma}~
  \label{eq:CMBCriterionpre}
\end{eqnarray}
where 
\begin{equation}
      N_{\mathrm{CMB}} ~\equiv~ \left({m_0\over \Delta m}\right)^{1/\delta}
   \left[\left(\frac{\tau_0}{t_{\mathrm{CMB}}}\right)^{\gamma}-1\right]^{1/\delta} 
\end{equation}
is the highest value of the index $n$ for which $\phi_n$ has a lifetime longer than 
a fiducial early time $t_{\mathrm{CMB}} \sim 10^{11}$~s.  This is approximately the time
before which decays have little effect on the CMB.~  
Note that since $\alpha+\gamma<0$,
the individual constituent contributions to $\xi$ necessarily decrease 
as a function of increasing mass within the ensemble.
This helps soften the sensitivity of $\xi$ to the precise value of $N_{\rm CMB}$.

In Fig.~\ref{fig:CMBCriterion}, we show contours of the energy-injection 
parameter $\xi$ in the $(\tau_0,\alpha + \gamma)$-plane for $\alpha = -3$ (left 
panel) and $\alpha = -2$ (right panel).  For both panels, we have set 
$\Delta m = 1$~GeV, $\delta = 1$, and $m_0 = 500$~GeV, although we emphasize
that the results shown here are essentially insensitive to the choice of 
$m_0$ within the range $100\mathrm{~GeV} \lesssim m_0 \lesssim 1500\mathrm{~GeV}$.~  
The hatched region demarcated 
by the solid black curve is the region of parameter space excluded by the 
CMB-consistency criterion in Eq.~(\ref{eq:CMBCriterionGeneralDDM}).  
It is clear from the figure that our CMB criterion imposes a
non-trivial constraint on the parameter space of our DDM model.  Indeed, for 
the range of values of $\tau_0$ relevant for reproducing the observed positron excess,
consistency with this criterion essentially requires $\alpha + \gamma \lesssim -1$ and a 
lifetime $\tau_0 \gtrsim 10^{25}$~s.  

\subsection{Combined electron/positron flux constraints}

Finally, we require that the combined $e^\pm$ flux from the DDM ensemble agree
with the combined $e^\pm$ flux reported by the FERMI 
collaboration~\cite{FERMIPositronData2010} to within $3\sigma$.  Note that 
we evaluate the goodness of fit for this combined flux in the same manner as 
for the gamma-ray flux, except that the corresponding $\chi^2$ 
statistic does not include the Heaviside-theta-function factor which appears 
in Eq.~(\ref{eq:ChiSqForFERMI}).


\section{Results\label{sec:results}}


Having outlined the phenomenological constraints that we require our DDM model 
to satisfy, we now turn to the main issue of this paper:  to what extent can 
we construct DDM models of the sort outlined in Sect.~\ref{sec:ensembles} which 
not only satisfy these constraints but also agree with the recent data from the 
\mbox{AMS-02} experiment concerning the positron fraction for energies up to 
$E_{e^+}\approx 350$~GeV?  And even more importantly, to what extent can we 
then {\it predict}\/ the behavior of the positron fraction 
for even higher energies, in the range  
$350~{\rm GeV}\lsim E_{e^+} \lsim 1~{\rm TeV}$? 

In order to address these questions, we adopt the following procedure.  First, 
as discussed in Sect.~\ref{sec:ensembles}, 
we survey over the parameter space $(\alpha,\gamma,m_0)$ of our DDM model,
fixing $\Delta m = 1$~GeV and $\delta = 1$. 
For each point 
$(\alpha,\gamma,m_0)$ in the DDM parameter space,
we then perform a best-fit analysis for the lifetime $\tau_0$ of 
the lightest mode as well as for the overall normalization factor $k$ associated with
the background electron flux (restricted to the range $0.7 \leq k \leq 1.0$).
Finally, we 
determine the minimum statistical significance with which the corresponding 
ensemble reproduces the results obtained by the \mbox{AMS-02} experiment within the
energy range $20\mathrm{~GeV} < E_{e^\pm} < 350\mathrm{~GeV}$
while simultaneously satisfying all of our consistency criteria.
Once again, as with the combined $e^\pm$ flux,
we evaluate the goodness of fit for the positron fraction in the same manner 
as for the gamma-ray flux, except that the corresponding $\chi^2$ 
statistic does not include the Heaviside-theta-function factor 
which appears in Eq.~(\ref{eq:ChiSqForFERMI}).

Our results are as follows.  For a DDM ensemble whose constituents $\phi_n$ 
are bosonic and decay either primarily to $e^+e^-$ 
or primarily to $\tau^+\tau^-$, we find no combination of parameters for 
which our consistency criteria are satisfied and the ensemble simultaneously
yields a positron-fraction curve which accords with \mbox{AMS-02} results within 
$5\sigma$.  In particular, we find that the $\phi_n \rightarrow \tau^+\tau^-$ 
channel tends to overproduce gamma rays while the $\phi_n \rightarrow e^+e^-$ 
channel tends to produce too hard an energy spectrum --- even within the 
context of a DDM ensemble.

\begin{figure*}
\begin{center}
  \epsfxsize 2.75 truein \epsfbox {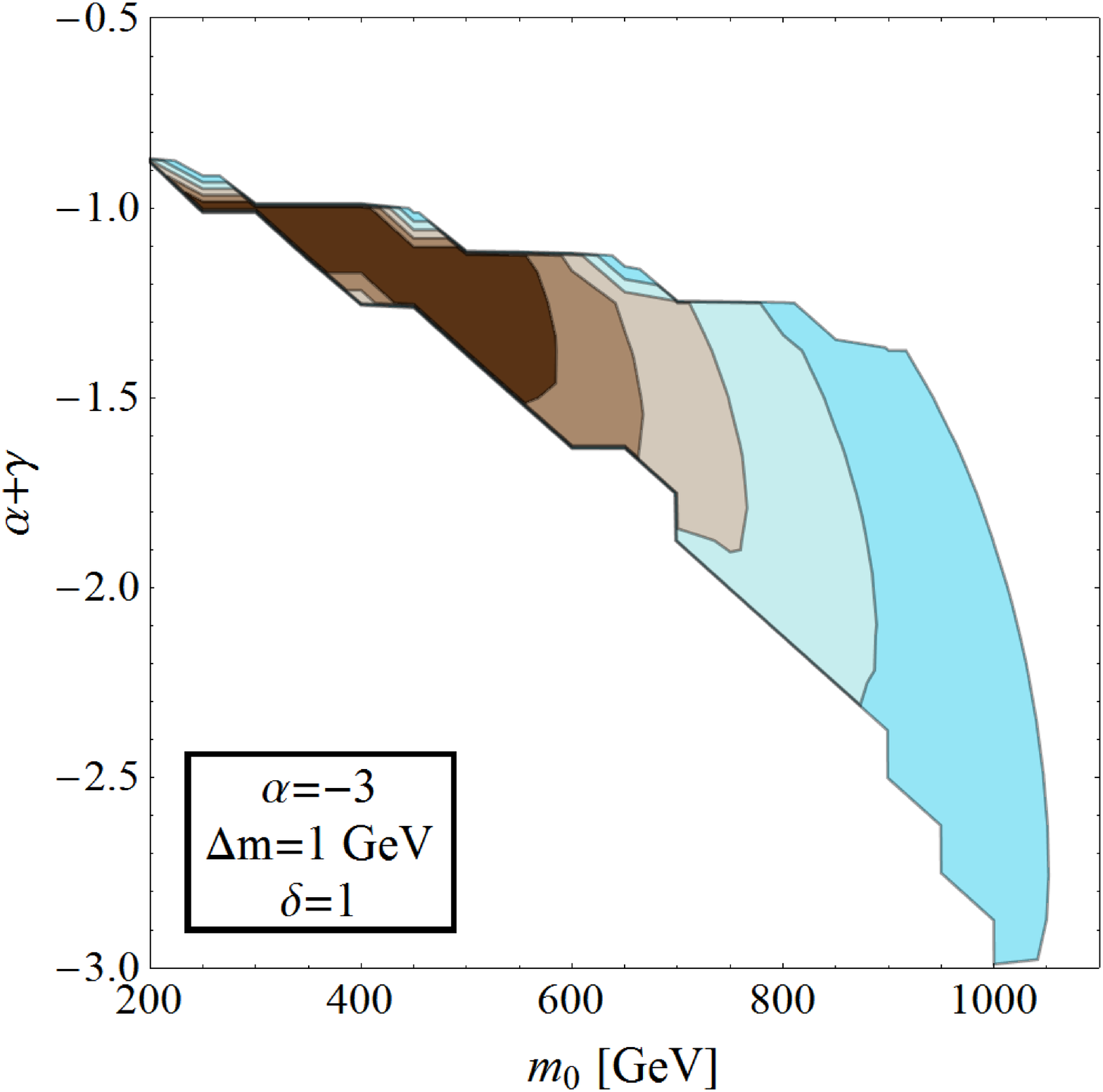}
  \hskip 0.4 truein
  \epsfxsize 2.75 truein \epsfbox {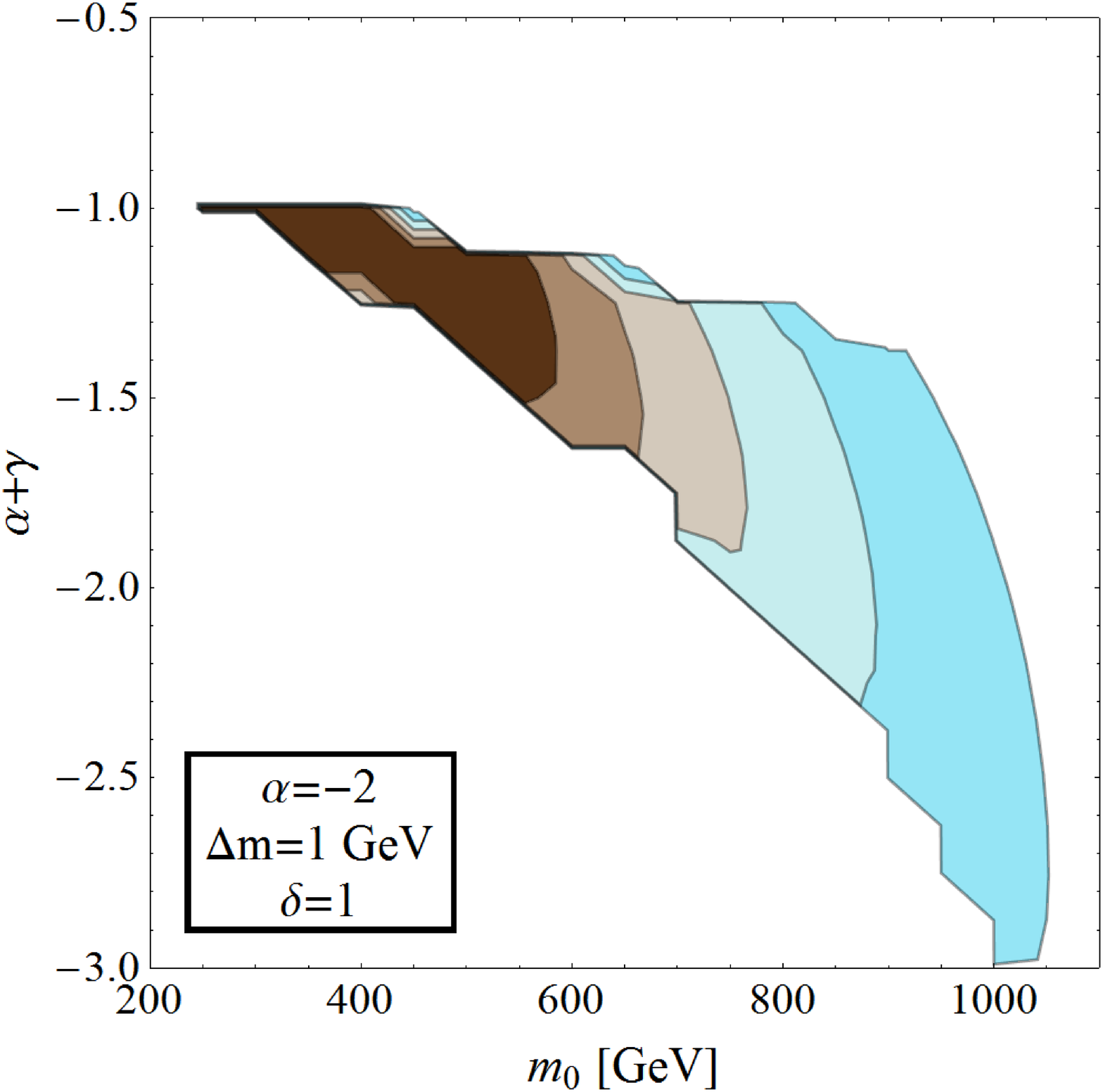} \\
  \raisebox{0.5cm}{\large Significance:~~}
    \epsfxsize 5.00 truein \epsfbox {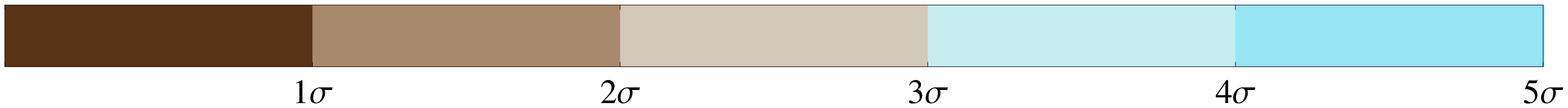}
\end{center}
\caption{Contours 
of the minimum significance level with which
a given DDM ensemble is consistent with \mbox{AMS-02} data, plotted within
the $(m_0,\alpha + \gamma)$ DDM parameter space 
for $\alpha= -3$ (left panel) and $\alpha= -2$ (right panel).
The colored regions correspond to DDM ensembles which successfully reproduce the \mbox{AMS-02} data 
while simultaneously satisfying all of the applicable
phenomenological constraints outlined in Sect.~\ref{sec:constraints},
while the 
white regions of parameter space 
correspond to DDM ensembles
which either cannot simultaneously satisfy these constraints
or which fail to match the \mbox{AMS-02} positron-excess data at the $5\sigma$ 
significance level or greater.
The slight difference between the results 
shown in the two panels is a consequence of the differences in the CMB constraints for the 
two corresponding values of $\alpha$.}  
\label{fig:SignificanceContourPlot} 
\vskip 0.2 truein
\begin{center}
  \epsfxsize 7.0 truein \epsfbox {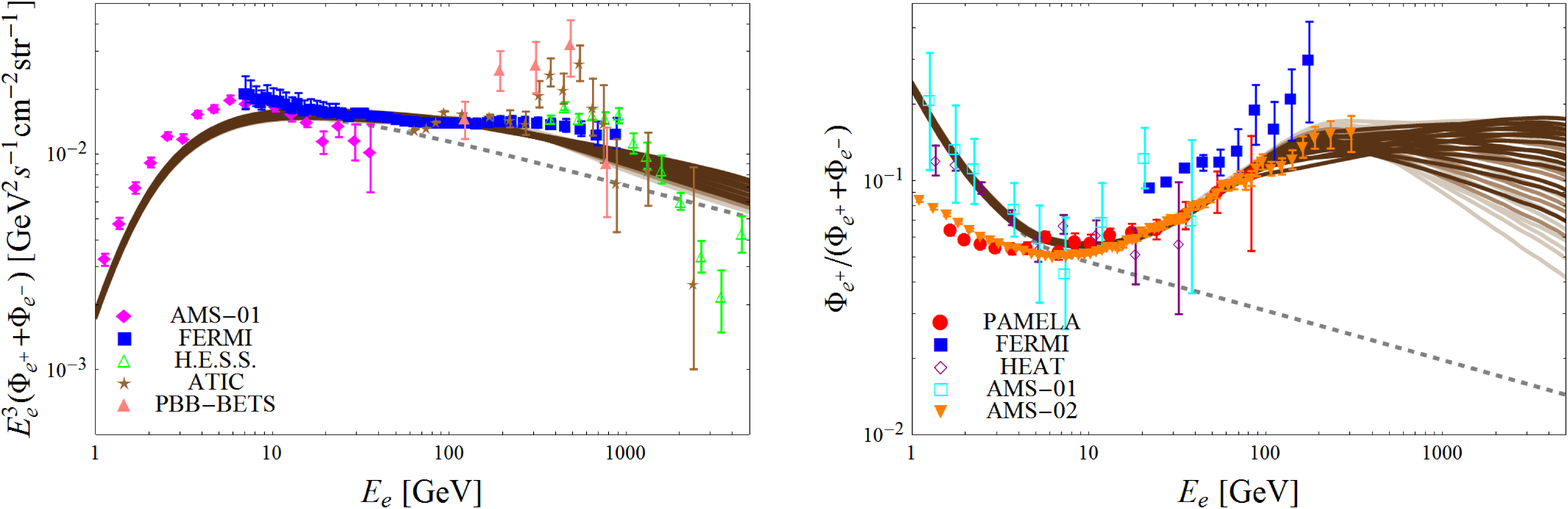}
\end{center}
\caption{Predicted combined fluxes $\Phi_{e^+} + \Phi_{e^-}$ (left panel)
and positron fractions $\Phi_{e+}/(\Phi_{e^+} + \Phi_{e^-})$ (right panel)
corresponding to the
DDM parameter choices lying within
those regions of Fig.~\ref{fig:SignificanceContourPlot} 
for which our curves agree with \mbox{AMS-02} data to within $3\sigma$. 
These curves are therefore all consistent 
with current combined-flux data to within $3\sigma$ and also 
consistent with current positron-fraction data to within $3\sigma$
(with the color of the curve indicating the precise quality of fit, using
the same color scheme in Fig.~\protect\ref{fig:SignificanceContourPlot}).   
These curves are also
consistent with all other applicable phenomenological constraints discussed in Sect.~\ref{sec:constraints}.  
However, despite these constraints, the behavior of the positron-fraction 
curves beyond $E_{e^\pm}\sim 350$~GeV is 
entirely unconstrained except by the internal theoretical structure
of the DDM ensemble.  Their relatively flat shape in this energy range
thus serves as a prediction (and indeed a ``smoking gun'') of the DDM framework.
Data from \mbox{AMS-02}~\cite{AMS02DataNew}, HEAT~\cite{HEATData1997PosFrac},
AMS-01~\cite{AMS01Data2002}, PAMELA~\cite{PAMELAData2010PosFrac}, 
FERMI~\cite{FERMIPositronData2011,FERMIPositronData2010}, PBB-BETS~\cite{PBBBETSData}, 
ATIC~\cite{ATICData}, and HESS~\cite{HESSData} are also shown for reference.}
\label{fig:FanPlot} 
\end{figure*}

By contrast, for an ensemble whose constituents decay primarily 
to $\mu^+\mu^-$, we find that there exist large regions of parameter space 
within which all of our criteria are satisfied and within which the DDM ensemble 
provides a good fit to \mbox{AMS-02} data.  
In Fig.~\ref{fig:SignificanceContourPlot}, 
we indicate these regions of $(m_0,\alpha + \gamma)$ space by shading them 
according to the significance level within which the decaying DDM ensemble 
is capable of reproducing the positron-fraction results from \mbox{AMS-02}.  The 
results in the left and right panels correspond to the choices $\alpha = -3$ 
and $\alpha= -2$ respectively.
The white regions, by contrast, indicate those regions of parameter space within 
which our consistency criteria cannot simultaneously 
be satisfied while at the same time yielding a positron fraction which agrees 
with \mbox{AMS-02} at the $5\sigma$ significance level or better.
Note that the difference between the results shown in the two panels is extremely
slight, and is due to a slight weakening of the CMB constraint with decreasing
$\alpha$.

We see, then, that a DDM ensemble whose constituent particles decay primarily to 
$\mu^+\mu^-$ can indeed account for the observed positron excess while at the same 
time satisfying other phenomenological constraints on decaying dark matter.  
The underlying reason for this success is easy to understand
upon comparison with the case of a traditional dark-matter candidate
with the same decay phenomenology.
In a nutshell,
the $e^\pm$ injection spectra associated with traditional dark-matter 
candidates with the same decay phenomenologies are generally too hard,
and thus cannot match the softer \mbox{AMS-02} data after propagation through the ISM.
 ~By contrast, in the DDM framework, the total dark-matter cosmological 
abundance $\OmegaDM$ is partitioned across an ensemble of individual constituents 
with different masses.  This in turn leads to a softening of the resulting $e^\pm$ 
injection spectra.  
Furthermore, a traditional dark-matter candidate must generally 
be quite heavy in order to reproduce the observed positron fraction, with a mass 
$m_\chi \gtrsim 1$~TeV.~  For such a heavy dark-matter candidate, it is difficult 
at the same time to reproduce the combined FERMI flux $\Phi_{e^+} + \Phi_{e^-}$; 
moreover, for such candidates, constraints related to the gamma-ray flux 
are quite severe.  However, we see from 
Fig.~\ref{fig:SignificanceContourPlot} that the preferred region of parameter space 
for our DDM model is one in which a significant fraction of the dark-matter cosmological
abundance $\OmegaDM$ is carried by constituents with masses in the range 
$200\mathrm{~GeV} \lesssim m_n \lesssim 800\mathrm{~GeV}$.  For such light 
particles, the gamma-ray constraints are less severe.

It is this observation which lies at the heart of the phenomenological success of 
the DDM ensemble.  Moreover, according to the results of Ref.~\cite{DDMDirectDet},
the preferred regions of DDM parameter space indicated in 
Fig.~\ref{fig:SignificanceContourPlot} correspond directly to those regions in 
which the full DDM ensemble contributes meaningfully to the cosmological 
dark-matter abundance $\OmegaDM$ (as opposed to regions in which the single 
most-abundant constituent effectively carries the entirety of $\OmegaDM$).  
Thus, we see that it is the full set of degrees of freedom within the DDM 
ensemble which play a role in achieving this outcome.

Having demonstrated that DDM ensembles can successfully reproduce the positron 
fraction reported by \mbox{AMS-02}, we now turn to the all-important question of how the 
predicted DDM positron fraction behaves at energies $E_e > 350$~GeV.~  In this 
way, we will not only be probing the phenomenological predictions of the DDM 
framework, but also be determining the extent to which further data from \mbox{AMS-02} 
or from other forthcoming cosmic-ray experiments might serve to distinguish 
between DDM ensembles and other explanations of the positron excess.  
          
Our results are shown in Fig.~\ref{fig:FanPlot}.
In this figure, superimposed on the actual experimental data, we have 
plotted the predicted combined flux and positron fraction 
which correspond to a variety of DDM parameter  
choices that lie within those regions of Fig.~\ref{fig:SignificanceContourPlot} 
for which our curves agree with \mbox{AMS-02} data to within $3\sigma$. 
The color of each such curve reflects the 
significance level to which the predicted and observed positron fractions agree, 
using the same color scheme as in Fig.~\ref{fig:SignificanceContourPlot}.  
We emphasize once again that the 
values of $\tau_0$ and $k$ for each curve shown in Fig.~\ref{fig:FanPlot}   
are those for which the best fit to the positron fraction is obtained, irrespective 
of the goodness of fit to the combined $e^\pm$ flux, provided that this goodness of 
fit corresponds to a statistical significance of at most $3\sigma$.  As a result, 
the curves shown in the left panel of this figure essentially all deviate from FERMI 
data at the $3\sigma$ significance level.  However, 
substantially improved consistency with FERMI 
data can easily be achieved without significantly sacrificing consistency with 
\mbox{AMS-02} data --- \eg,
by employing an alternative fitting procedure involving a combined fit 
to both data sets simultaneously.             
Note also that the fit to FERMI data depends on a number of assumptions
concerning the astrophysical background flux, and not merely its
normalization;  hence small deviations from FERMI $\Phi_{e^+} + \Phi_{e^-}$
results are not necessarily to be taken as a sign of tension with data.

We immediately see from Fig.~\ref{fig:FanPlot} that DDM ensembles give rise to 
unusual and distinctive positron-fraction curves whose behaviors at high energies
differ significantly from those obtained for traditional dark-matter models.  
Indeed, traditional dark-matter models for explaining the observed positron excess predict 
a rather pronounced downturn at $E_{e^{\pm}} \lesssim m_\chi$ (for annihilating 
dark matter) or $E_{e^{\pm}} \lesssim m_\chi/2$ (for decaying dark matter),
where $m_\chi$ denotes the mass of the dark-matter particle.  By contrast, as we 
see in Fig.~\ref{fig:FanPlot}, DDM ensembles give rise to positron-fraction curves 
which either decline only gradually or remain effectively flat for 
$E_{e^\pm} \gsim 350$~GeV.~  Indeed, we see that 
$\Phi_{e^+}/(\Phi_{e^+} + \Phi_{e^-})\lesssim 0.2$ over this entire range.  
In principle, of course, DDM ensembles can give rise to positron-fraction 
curves exhibiting a broad variety of shapes and features.  However, imposing 
the requirements that the positron fraction and combined $e^\pm$ flux agree with 
current data substantially limits the high-energy behaviors for the resulting 
positron fraction, permitting only those curves for which this fraction levels 
off and remains relatively constant as a function of energy.  

Since no sharp downturn in the positron fraction appears consistent within the 
DDM framework, we may take this to be an actual prediction of the framework.
The presence or absence of such a downturn therefore offers a powerful tool 
for distinguishing decaying DDM ensembles from other dark-matter explanations 
of the positron excess.  Even more importantly, however,
we observe that a positron fraction which falls only
gradually or which remains effectively constant for $E_{e^\pm} \gsim 350$~GeV
can be achieved only in a scenario in which an ensemble of dark-matter states
with different masses and decay widths act together in coordinated fashion
in order to support the positron-fraction function against collapse and 
to carry it smoothly into this higher energy range.
This behavior, if eventually observed experimentally,
can therefore be taken as a virtual ``smoking gun'' of Dynamical Dark Matter.

This claim, of course, rests upon the fundamental assumption that we are 
attributing the positron excess to dark-matter physics. As we have 
indicated above, it is always possible that some configuration of pulsars 
or other astrophysical sources can also provide part or all of the 
explanation for the observed positron excess. Given this observation, it 
may initially seem that our conclusions regarding the spectra predicted by 
the DDM framework may be somewhat moot. However, the success of the DDM 
framework not only in accommodating the existing positron data but also in 
predicting the continuation of the positron excess out to 1~TeV can be taken, 
conversely, as indicating that one need not be forced into a conclusion 
involving traditional astrophysical sources should such phenomena 
be observed experimentally. Indeed, as we have shown, there 
exists a well-motivated dark-matter framework --- namely that involving a 
DDM ensemble obeying well-ordered scaling relations --- which can easily 
do the job.


\section{Varying the Input Assumptions\label{sec:varyinginputs}}


In the previous section, we showed that the DDM framework can naturally accommodate
the positron excess, that this framework actually predicts that the positron excess will continue
out to energies of at least approximately $E\sim 1$~TeV,  and that
any future experimental verification
of this prediction can actually be taken
(within the confines of dark-matter interpretations of the positron excess) 
as a ``smoking gun'' for DDM.~
However, there are a number of input assumptions which have either explicitly or implicitly played a role
in our analysis --- some of these concern the structure of the DDM ensemble itself, while others
concern the background astrophysical environment whose properties also enter into our calculations.
It is therefore critical that we understand the extent to which our results are 
robust against variations in these input assumptions.

\subsection{Varying the structure of the DDM ensemble}

As discussed in Sect.~\ref{sec:ensembles}, our DDM ensemble can be parametrized
in terms of five fundamental parameters $\lbrace \alpha,\gamma,\delta, m_0,\Delta m\rbrace$.
(The remaining parameters $\lbrace \Gamma_0,\Omega_0\rbrace$ are then essentially determined
through the ``normalization'' conditions discussed earlier.)
In the analysis we have presented thus far, we have allowed $\alpha$, $\gamma$, and $m_0$ to vary,
but we have taken $\delta=1$ and $\Delta m = 1$~GeV as fixed benchmarks.
Since there is nothing in the DDM framework which requires these particular values,
it is reasonable to ask what new effects might emerge if these values are altered.

\begin{figure*}
\begin{center}
  \epsfxsize 7.0 truein \epsfbox {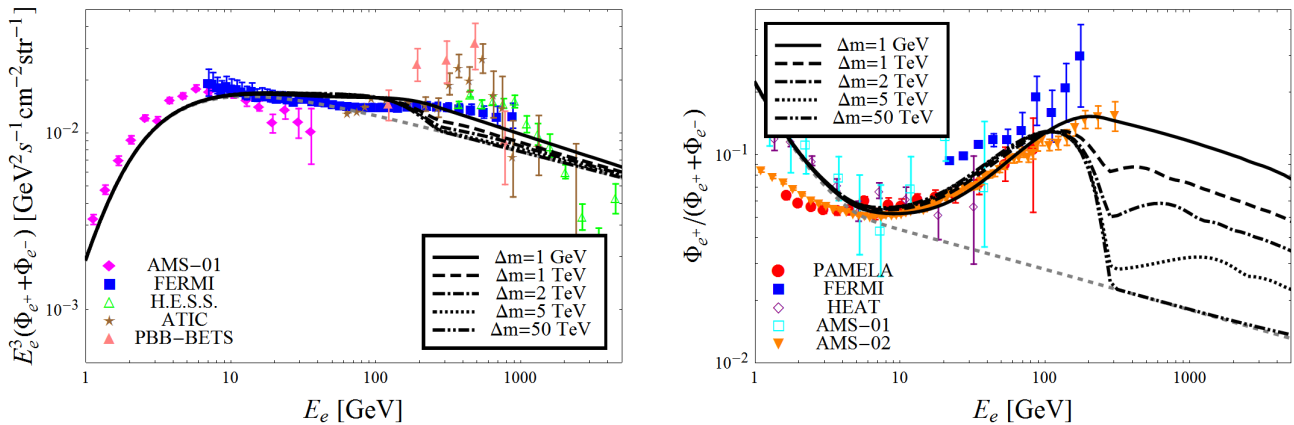}
\end{center}
\caption{The combined flux $\Phi_{e^+} + \Phi_{e^-}$ (left panel)
and positron fraction $\Phi_{e+}/(\Phi_{e^+} + \Phi_{e^-})$ (right panel)
corresponding to DDM parameter choices 
$\Delta m= \lbrace 10^{-3}, 1, 2, 5, 50\rbrace$~TeV.~
For these plots we have held
the other DDM parameters fixed
at reference values  
$\alpha = -2$, $\gamma= 0.5$, $\delta =1$,
$m_0= 600$~GeV, and 
$\tau_0 = 3.23\times 10^{26}$~s,
and we have taken $k=0.9$ in Eq.~(\protect\ref{eq:BGFluxesFunctionalForms}).
As expected, passing to larger values of $\Delta m$ has the effect of decreasing
the flux contributions from heavier states in the DDM ensemble;  the resulting
fluxes thus increasingly fail to match the existing highest-energy \mbox{AMS-02} data
for the positron fraction and at the same time 
begin to exhibit the characteristic downturn in the positron fraction
that is normally associated with traditional single-component dark-matter candidates.
Indeed, only by taking $\Delta m$ sufficiently small does the predicted positron fraction
match all of the \mbox{AMS-02} data.  However, a large portion of the 
DDM ensemble then plays an active role in contributing to these fluxes, whereupon
the internal structure of the ensemble itself compels the positron fraction to remain significantly
above background --- even out to energies $E_{e^\pm}\sim 1$~TeV --- without any sharp downturn.} 
\label{fig:deltamplot} 
\end{figure*}

The parameters $\Delta m$ and $\delta$ play 
independent but correlated roles:  both appear in Eq.~(\ref{eq:MassSpectrum}) and together
they parametrize the density of states in the DDM ensemble.
Increasing $\delta$ or $\Delta m$ has the effect
of increasing the masses of the heavier DDM constituents
relative to the lighter ones, thereby potentially diminishing their effects 
on low-energy physics.
Indeed, for sufficiently large $\Delta m$, our DDM ensemble essentially 
acts as a traditional single-particle
dark sector as far as most low-energy effects are concerned.
By contrast, decreasing $\delta$ or $\Delta m$ has the reverse effect.
Indeed, in the $\Delta m\to 0$ limit, the states in the DDM ensemble form a continuum,
and in practice such ``continuum'' behavior can be expected whenever
$\Delta m$ is smaller than the scale set by the energy resolution of the relevant 
cosmic-ray detectors.

In our analysis thus far, we were motivated to take \mbox{$\delta=1$} because
this is the value which arises in certain well-motivated realizations of DDM ensembles involving
large extra spacetime dimensions~\cite{DynamicalDM1,DynamicalDM2}.
However, our choice of $\Delta m=1$~GeV was made on purely aesthetic grounds,
as a small value of this size ensures that a large portion of the DDM ensemble plays a role in contributing
to the relevant cosmic-ray fluxes. 
It is therefore interesting to understand the extent to which our predictions remain valid
even if $\Delta m$ exceeds this value.

In Fig.~\ref{fig:deltamplot}, we show the fluxes which emerge from a variety of 
DDM ensembles corresponding to different values of $\Delta m$, while the remaining
DDM parameters are held fixed at values which ensure a successful fit to \mbox{AMS-02} data
for $\Delta m=1$~GeV.~
As we see from this figure, increasing the value of $\Delta m$ 
causes the resulting fluxes to increasingly deviate from
the existing \mbox{AMS-02} data, particularly at the highest energies
for which such data is available.
This is particularly dramatic for the positron-fraction data;
indeed, increasing the value of $\Delta m$ ultimately reintroduces  
the characteristic downturn 
that is normally associated with traditional single-component dark-matter candidates.
As we see, it is only by taking $\Delta m$ sufficiently small that the predicted positron fraction
can match all of the \mbox{AMS-02} data.  This is precisely the ``DDM limit'' in which a large
portion of the DDM ensemble plays an active role in contributing to the cosmic-ray fluxes
at these energies.
     
This observation demonstrates that there is a strong correlation between obtaining a successful 
fit to the \mbox{AMS-02} data and having a dark sector such as a DDM ensemble in which
a relatively large number of dark-matter states actually contribute to the cosmic-ray fluxes at these energies.
But what is particularly remarkable about the results shown in Fig.~\ref{fig:deltamplot} is
the existence of a second, independent correlation:  a successful fit to the entirety of available \mbox{AMS-02}
data is also correlated with the absence of a sharp downturn in the positron fraction
out to energies $E_{e^\pm}\sim 1$~TeV.~
Indeed, setting $\Delta m$ to any value less than the maximum value 
that will fit the existing \mbox{AMS-02} data 
causes the corresponding positron fraction to exhibit at most
a gently declining plateau out to $E_{e^\pm}\sim 1$~TeV.~
In an arbitrary hypothetical multi-component theory of dark matter, this second correlation need not have existed, 
but its emergence in the DDM framework is ultimately a consequence of the tight internal scaling structure 
of the DDM ensemble.
It is this ``rigidity'' of the DDM ensemble --- \ie, its inability to fit the existing \mbox{AMS-02} data while
simultaneously producing an immediate downturn in the positron fraction at higher energies --- 
which is the underlying reason that the DDM framework
yields such ``smoking gun'' predictions about the positron fraction at higher energies.
(In this context we remark that although we have taken $m_0= 600$~GeV for the curves in Fig.~\ref{fig:deltamplot},
altering $m_0$ does not change this conclusion and would in fact cause difficulties satisfying our other constraints
on the total gamma-ray flux or the total $\Phi_{e^+} + \Phi_{e^-}$  flux.) 
Thus, this correlation is relatively robust against variations in $\Delta m$, and we expect it to hold
for all values $\Delta m$ which are
sufficiently small as to permit a successful fit to the $\mbox{AMS-02}$ data.

In this connection, it is perhaps also worth commenting on the role 
played by the quantity $n_{\rm max}$ which truncates our sums over DDM constituents
in Eqs.~(\ref{eq:Omegatot}), (\ref{eq:DDMepemSourceTerm}), and so forth.
At first glance, it might seem that $n_{\rm max}$ is yet another free parameter in the DDM framework.
Even worse, if we attempt to 
interpret $n_{max}$ literally as the number of DDM states which contribute to the cosmic-ray
fluxes, we might be tempted to view any  
fit requiring a very large value of $n_{\rm max}$ 
as somehow uninteresting, since it might be expected on general grounds that
fits to data can always be performed 
with arbitrary precision if we have sufficiently many degrees of freedom at our disposal.
While these worries would certainly be valid for an arbitrary hypothetical 
multi-component theory of dark matter, the important point here is that 
our dark sector is not just a random collection of individual states with 
arbitrary, freely adjustable masses and decay widths.  Instead, these states are part of 
a DDM {\it ensemble}\/ which is collectively constrained by scaling relations
of the sorts described in Sect.~\ref{sec:ensembles}. 
Indeed, in the DDM framework, the dark sector is parametrized in terms of 
relatively few degrees of freedom (such as $\alpha$, $\gamma$, and $m_0$, as discussed above), 
and these quantities are chosen
in such a way as to eliminate any theoretical or numerical sensitivity
to the cutoff $n_{\rm max}$.    
At a practical level, this means that the cosmic-ray flux contributions from all DDM states
are completely and simultaneously fixed once these few parameters are specified,
and that successively heavier DDM states make increasingly smaller contributions
to these fluxes in such a way that all sums are convergent as $n_{\rm max}\to \infty$.

\begin{figure*}
\begin{center}
  \epsfxsize 2.1 truein \epsfbox {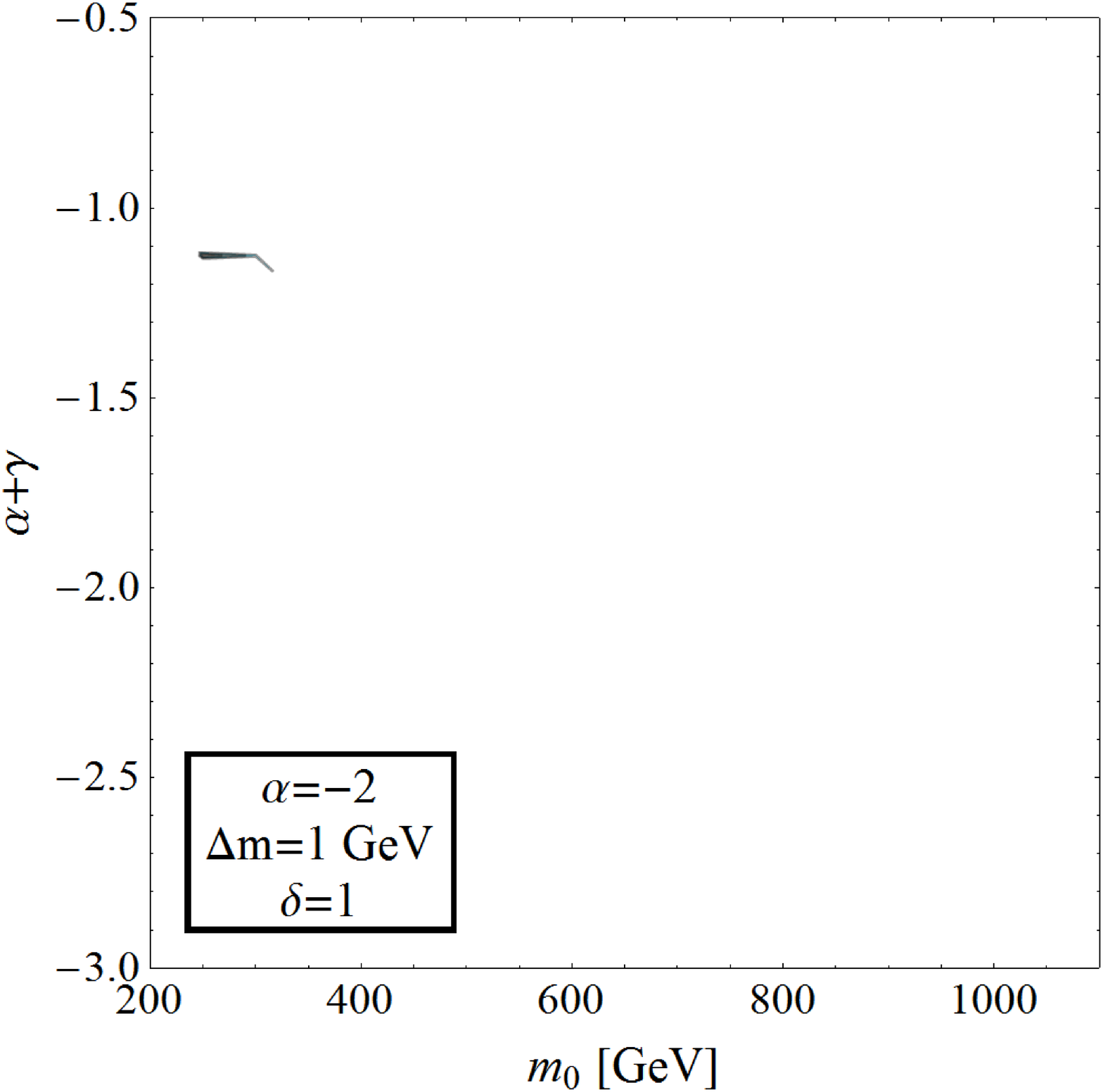} 
   \hskip 0.15 truein
  \epsfxsize 2.1 truein \epsfbox {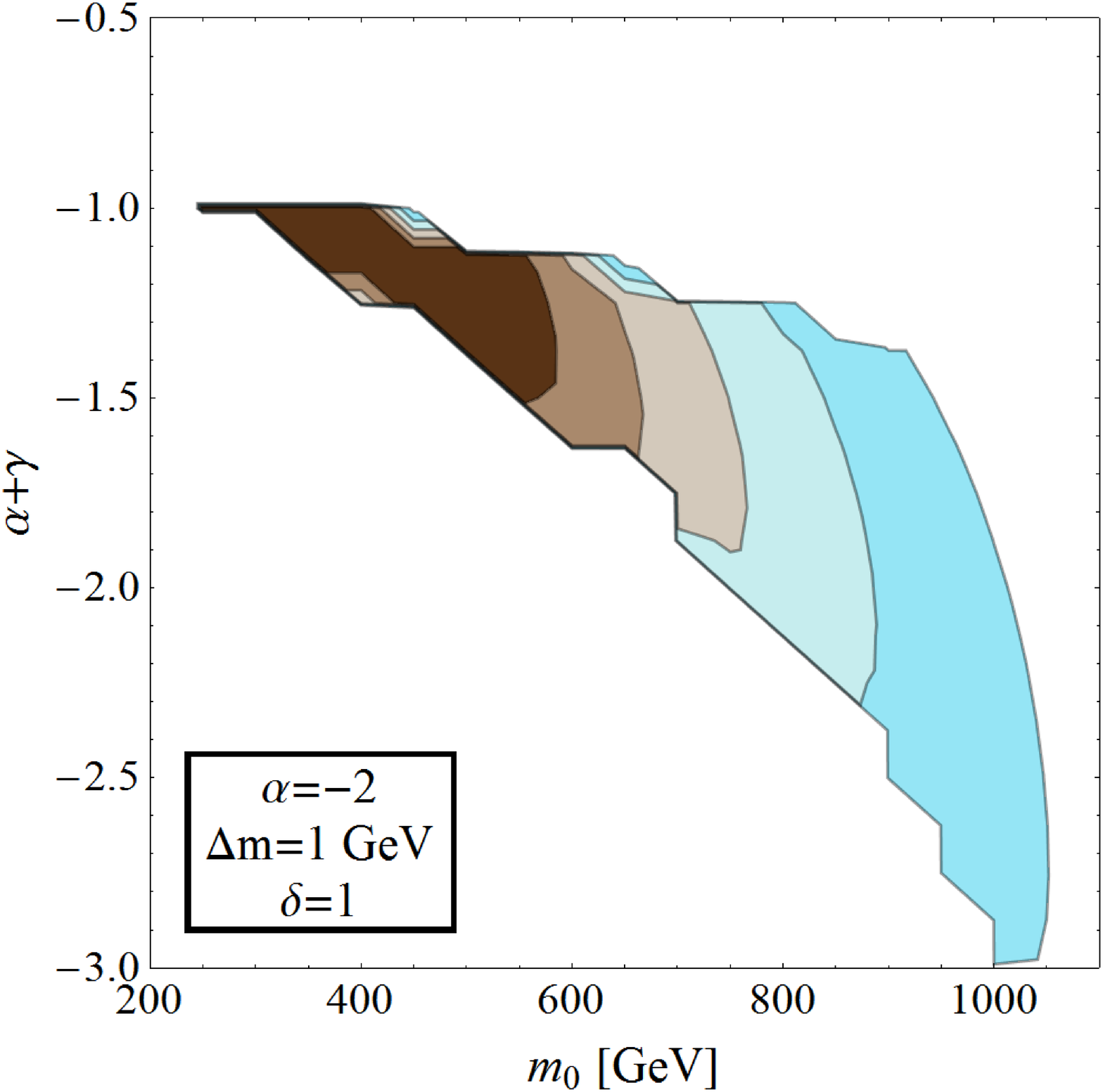}
  \hskip 0.15 truein
  \epsfxsize 2.1 truein \epsfbox {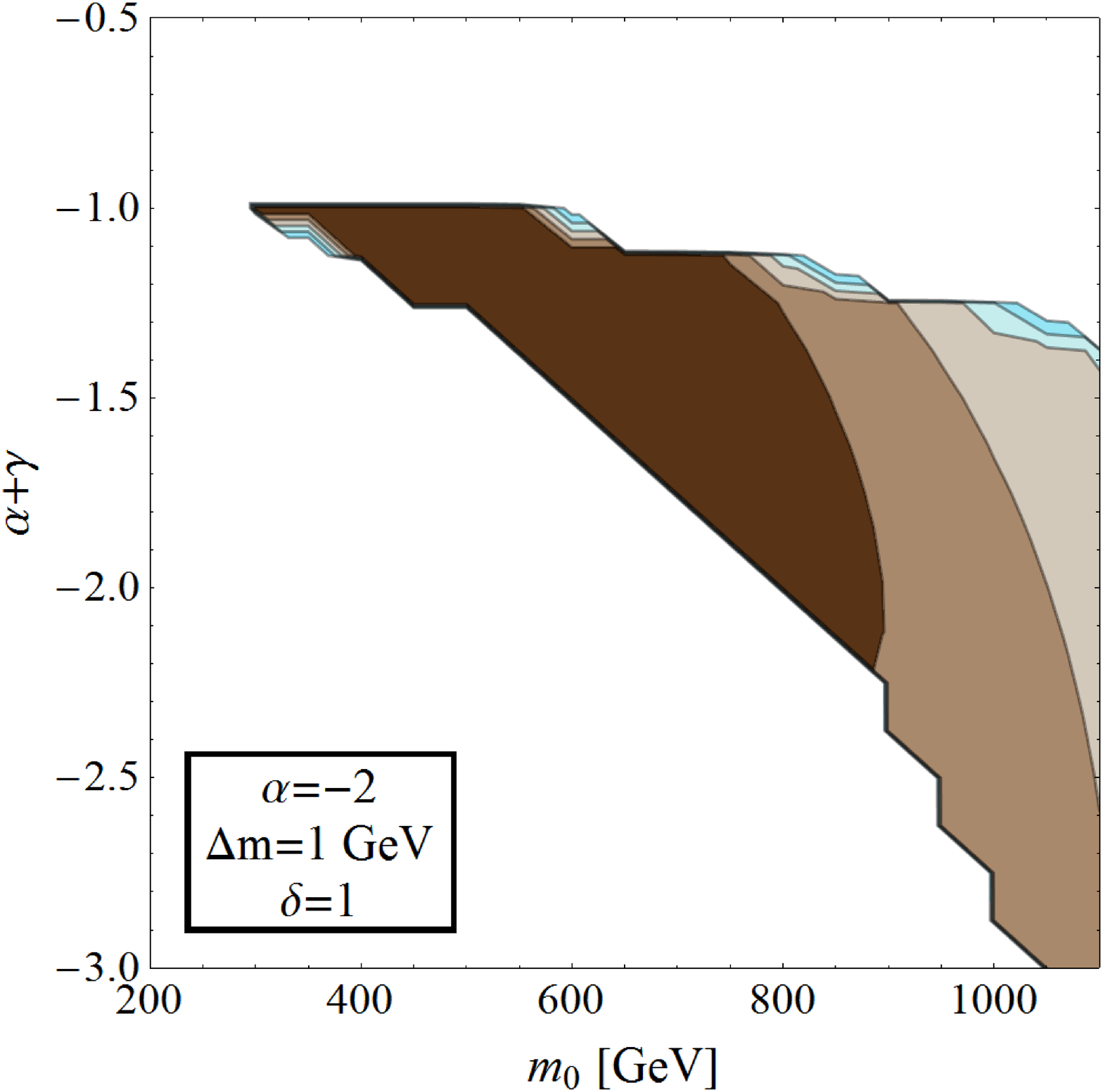} 
\end{center}
\begin{center}
  \epsfxsize 2.1 truein \epsfbox {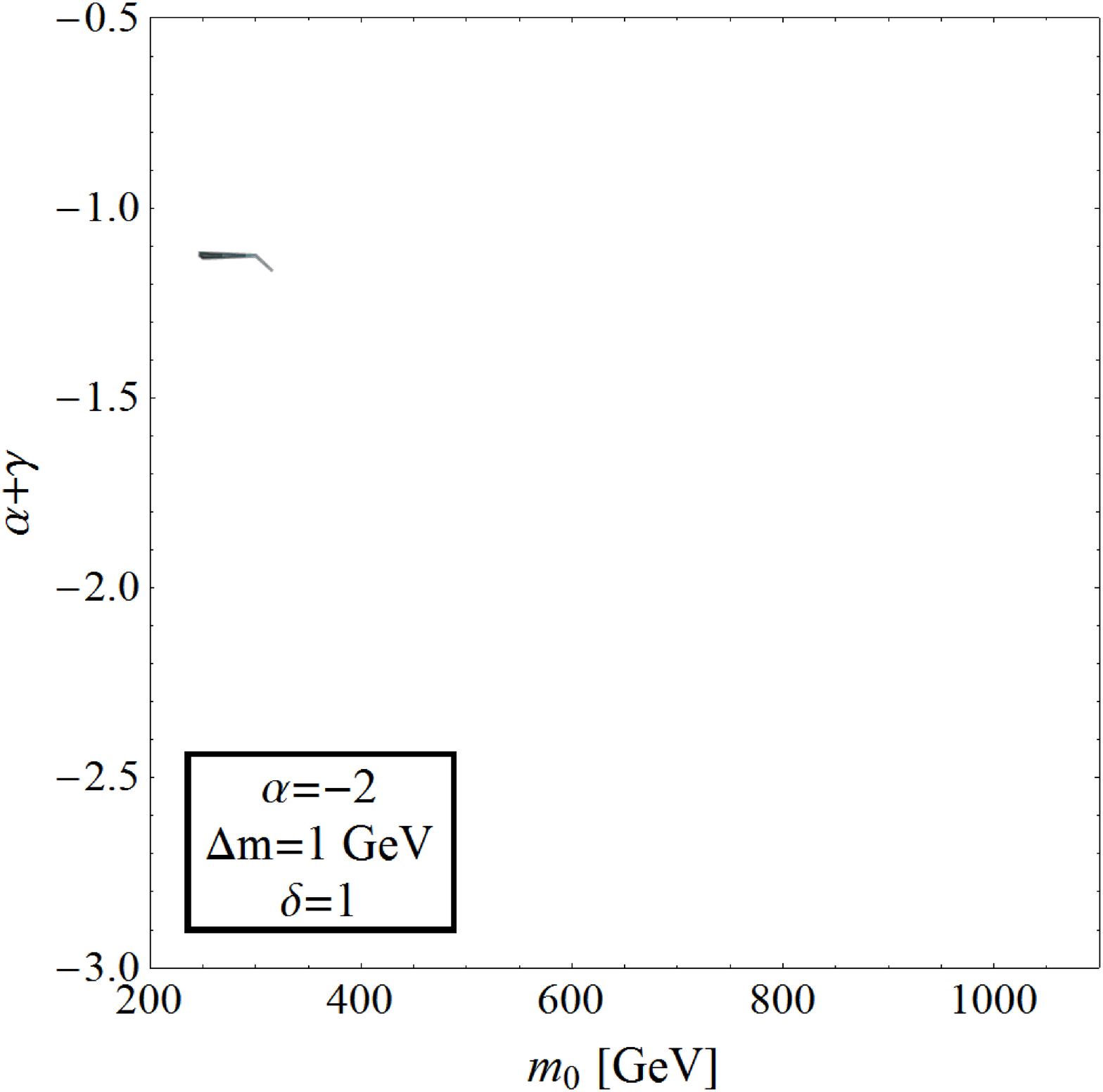} 
  \hskip 0.15 truein
  \epsfxsize 2.1 truein \epsfbox {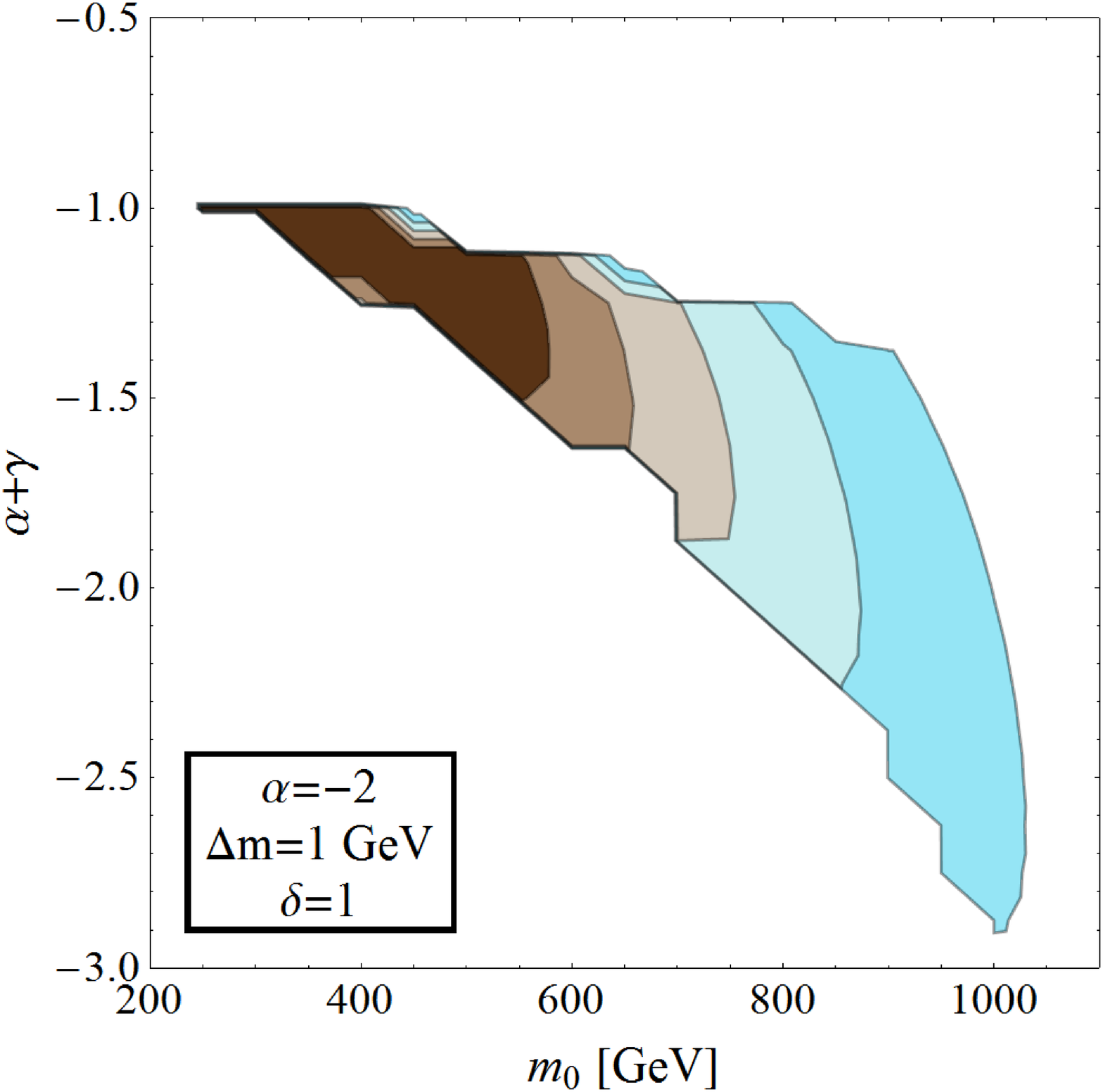}
  \hskip 0.15 truein
  \epsfxsize 2.1 truein \epsfbox {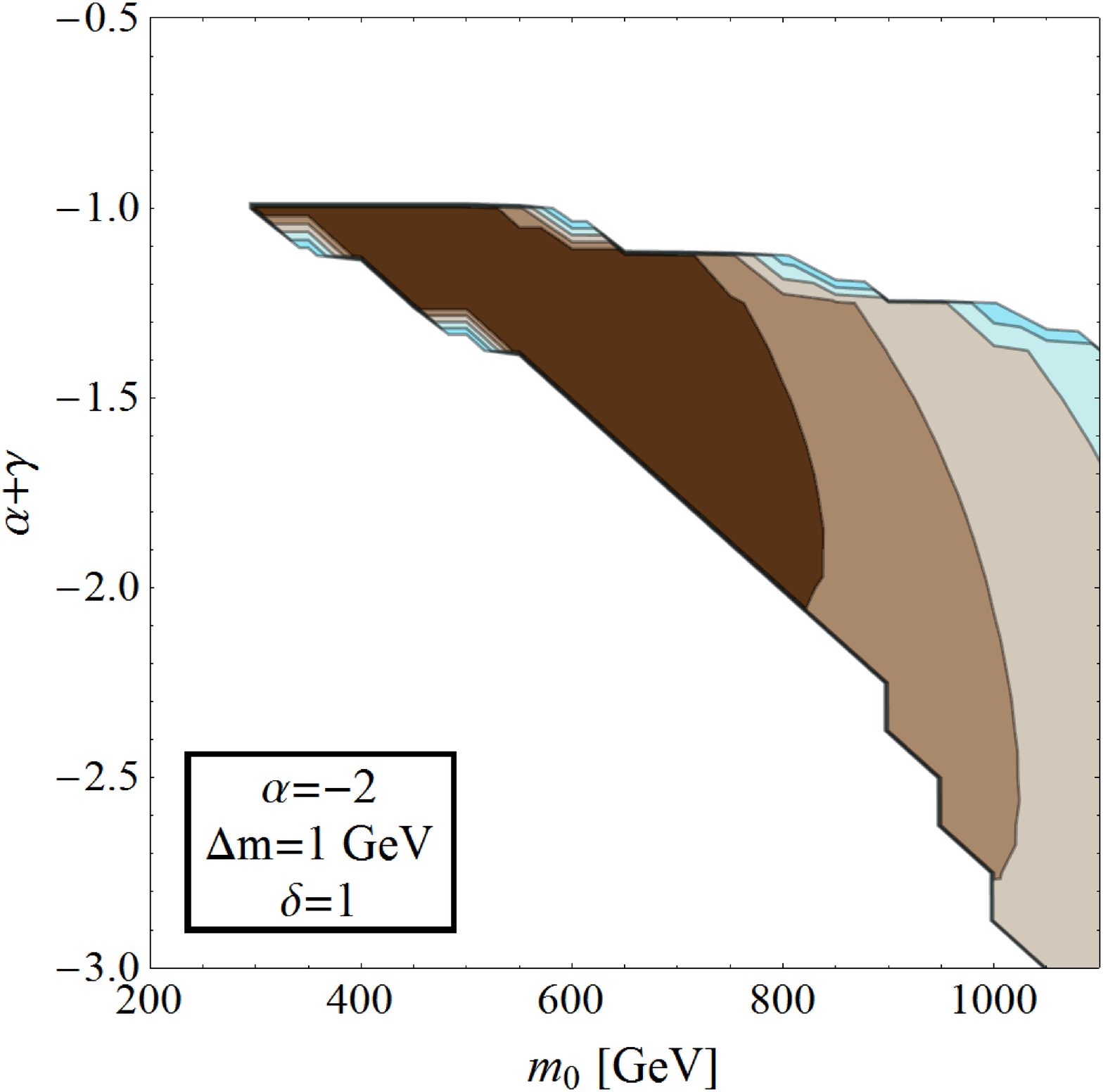}
\end{center}
\caption{Contours 
of the minimum significance level with which
a given DDM ensemble is consistent with \mbox{AMS-02} data, plotted 
exactly as in Fig.~\ref{fig:SignificanceContourPlot} but with $\alpha= -2$ only. 
The plots in the (left, middle, right) column 
are respectively calculated using the (MIN, MED, MAX) propagation model,
while those in the (top, bottom) row 
are respectively calculated assuming an (NFW, isothermal) dark-matter halo.
Thus the upper middle panel is identical to the right panel 
of Fig.~\ref{fig:SignificanceContourPlot},
and is reproduced here for comparison purposes.
We see that while the MIN propagation model results in a near-total elimination of 
the allowed DDM parameter space, the MED and MAX propagation models result
in allowed parameter spaces which are roughly equivalent, differing only in their
qualities of fit to the \mbox{AMS-02} data.  Likewise, we see
that our results are almost completely insensitive to the particular form of the
dark-matter halo assumed.}
\label{fig:varybackgroundsandfluxes} 
\end{figure*}

\subsection{Varying the astrophysical modelling}
 
Thus far we have concentrated on input assumptions 
associated with our DDM ensembles.
However, there were also a number of implicit assumptions of a purely
astrophysical nature.
In particular, in this category, 
two assumptions stand out:
\begin{itemize}
\item  We used a particular astrophysical propagation model (the so-called ``MED''
       propagation model) in order to calculate our DDM-produced cosmic-ray fluxes.
\item  We assumed a particular dark-matter halo profile (the so-called ``NFW'' profile)
       for our DDM ensemble.
\end{itemize}
As discussed in Sects.~\ref{sec:ensembles} and \ref{sec:propagation},
both of these assumptions were implicitly part of the flux calculations leading
to the results in Fig.~\ref{fig:SignificanceContourPlot}.
However, neither of these assumptions is required on theoretical grounds,
and in each case there exist alternative models which might have been chosen.
For example, rather than adopt the MED propagation model, we could have adopted its
siblings, the MIN or MAX propagation models~\cite{DelahayePositronPropagation,DonatoAntiprotonPropagation}.
Together, the MIN, MED, and MAX models 
reside within an entire class of propagation models which differ
in (and are therefore effectively parametrized in terms of)
the degree to which the input fluxes are ``processed'' (or
effectively shifted downwards in energy) in passing through the interstellar medium, with the
MIN (MAX) propagation model tending to minimize (maximize) the
resulting fluxes of charged cosmic-ray particles subject to certain phenomenological constraints.
Likewise, rather than adopt the NFW dark-matter halo, we could just as easily
adopted any of a number of other halo profiles.  For example, one well-motivated choice
might be the so-called ``isothermal'' dark-matter halo~\cite{isothermal} ---
this is nothing but the density distribution 
exhibited by an isothermal, self-gravitating system of particles,
and leads to a velocity dispersion which is essentially constant at large radii.
Note that in general, the isothermal dark-matter halo is generally smoother 
({\it i.e.}\/, less ``cuspy'') than the NFW halo at small radii and thus
forms a nice counterpoint to the NFW halo profile.

\begin{figure*}
\begin{center}
  \epsfxsize 7.0 truein \epsfbox {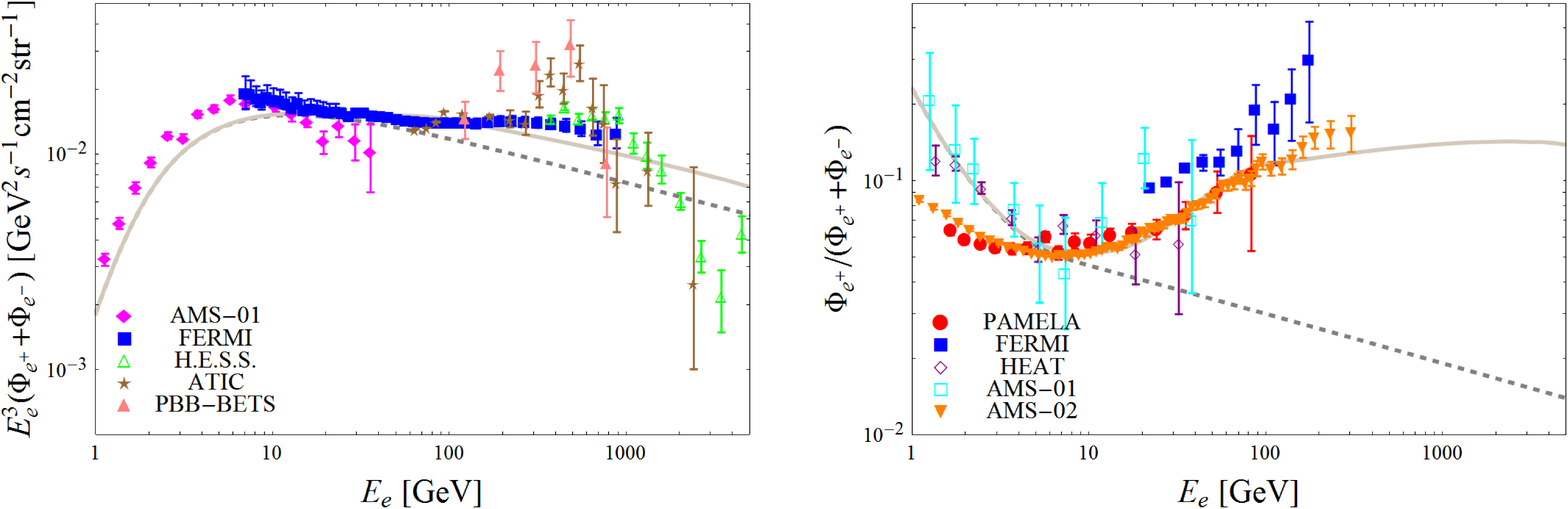}
\end{center}
\begin{center}
  \epsfxsize 7.0 truein \epsfbox {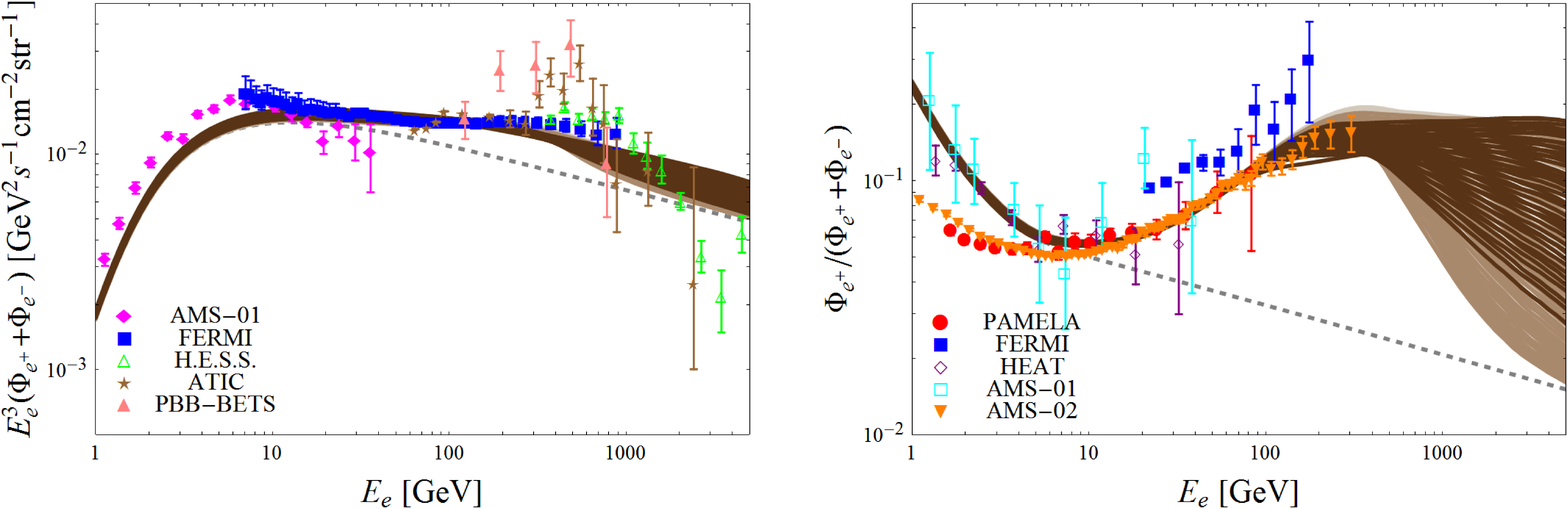}
\end{center}
\caption{The combined flux $\Phi_{e^+} + \Phi_{e^-}$ (left panels)
and positron fraction $\Phi_{e+}/(\Phi_{e^+} + \Phi_{e^-})$ (right panels)
corresponding to the DDM parameter choices 
illustrated in Fig.~\ref{fig:varybackgroundsandfluxes}
for the MIN propagation model (top row)
and the MAX propagation model (bottom row). 
Even though the MAX propagation model allows considerably greater
flexibility in terms of the behavior of the positron fraction
than the MIN or MED propagation models,
even this case does not permit a sudden, sharp downturn in the
positron fraction.  Indeed, the steepest allowed decline continues to be 
relatively slow, reaching the expected background flux only at 
energies beyond $E_{e^\pm}\sim 1$~TeV.}   
\label{fig:fanplotsMINMAX} 
\end{figure*}

In Fig.~\ref{fig:varybackgroundsandfluxes}, we show the degree to which
the results in Fig.~\ref{fig:SignificanceContourPlot} would be altered 
by such replacements.
As we see from Fig.~\ref{fig:varybackgroundsandfluxes},
the effect of replacing the MED propagation model with the MIN propagation
model is quite dramatic and results in a near-total elimination of the
allowed DDM parameter space, while replacement with the MAX propagation
model does not dramatically alter the allowed region of DDM parameter
space and merely changes the calculated quality of fit to \mbox{AMS-02} data
within this region.
These results are relatively easy to understand.   
Broadly speaking, the effect of propagation on the injected fluxes through the interstellar medium
is to increase the contribution to $\Phi_{e^\pm}(E_{e^\pm})$ at low 
$E_{e^\pm}$ from injected $e^\pm$ with an injection energy $E_{e^\pm}' > E_{e^\pm}$.
Moreover, this effect becomes increasingly pronounced at higher $E_{e^\pm}'$.  
The collective contribution to $\Phi_{e^\pm}$ from the heavier constituents in the DDM 
ensemble is therefore greater at low $E_{e^\pm}$ for propagation models such the MAX model, 
in which the processing of the $e^\pm$ injection spectra by the interstellar
medium is more pronounced.  As we have seen, 
this contribution is ultimately responsible for the all-important softening of the 
$\Phi_{e^\pm}$ flux spectra which renders these spectra compatible with the \mbox{AMS-02} data.  
Therefore, the increased processing of the injection spectrum by the interstellar medium
in the MAX 
model relative to the MED model can compensate to some degree for the hardening of 
the injection spectrum which results from an increase in $m_0$, as shown in Fig.~\ref{fig:deltamplot}.
By contrast, use of the MIN propagation model has the effect of minimizing the contributions to the
low-energy $e^\pm$ fluxes from the heavier DDM ensemble components.  The resulting near-total elimination of 
the allowed DDM parameter space therefore can be taken as yet another feature highlighting the critical
role of the entire DDM ensemble (and not just its lightest component) in producing a successful 
fit to $\mbox{AMS-02}$ data.

The other lesson that can be drawn from Fig.~\ref{fig:varybackgroundsandfluxes}
is that our results are not particularly sensitive to
the choice of halo profile.  This indicates that our results are indeed properties
of the DDM framework rather than properties of the specific 
halo profile chosen.  The primary reason for this halo-insensitivity is that the
dominant contribution to $\Phi_{e^+}$ and $\Phi_{e^-}$ in the DDM framework is
the result of dark-matter decay rather than dark-matter annihilation.
The contribution to $\Phi_{e^+}$ or $\Phi_{e^-}$ from a decaying
dark-matter particle $\phi$ is proportional to its density $\rho_\phi$, whereas 
for an annihilating particle it is proportional to $\rho_\phi^2$. 
These fluxes are consequently far less sensitive to the shape of the
dark-matter halo in the former case than in the latter.  Likewise, the
dominant constraints on decaying dark-matter models of the positron
excess come from considerations which depend either very weakly (or not at
all) on the shape of the galactic dark-matter halo, such as the diffuse
extragalactic gamma-ray flux and the properties of the CMB radiation.

In Fig.~\ref{fig:fanplotsMINMAX}, we show the fluxes which result   
from the use of the MIN and MAX propagation models within the corresponding
allowed parameter spaces shown in Fig.~\ref{fig:varybackgroundsandfluxes}.
As we saw in Fig.~\ref{fig:varybackgroundsandfluxes},
use of the MIN propagation model results in only a small surviving sliver
of DDM parameter space;  this in turn leads to a fairly sharp set of
flux predictions in Fig.~\ref{fig:fanplotsMINMAX}.  
By contrast, use of the MAX propagation model leads to a set of possible
fluxes in Fig.~\ref{fig:fanplotsMINMAX} which are even somewhat broader in their
allowed behaviors than those which appear in Fig.~\ref{fig:FanPlot} for the MED propagation model. 
Despite these differences, however, 
we once again see that the primary prediction of the DDM framework ---  
that the positron fraction will continue to exhibit a surplus out to the higher energies
$E_{e^+}\sim 1$~TeV without exhibiting a sharp downturn --- continues to hold in all cases.
Indeed, even the steepest allowed decline exhibited in the MAX case continues to be 
a relatively slow one in which the positron flux reaches its expected background value only at 
energies well beyond $E_{e^\pm}\sim 1$~TeV.   

We see, then, that the primary results we have presented in this 
paper --- that decaying DDM ensembles can successfully reproduce both \mbox{AMS-02} 
and FERMI data and yield concrete predictions for the positron fraction at high 
energies --- in most cases do not depend significantly on our choice of astrophysical 
approximations or computational tools.
Indeed, as discussed in Sect.~\ref{sec:results},
it is the relative softness of the $\Phi_{e^\pm}$ spectra from an entire decaying DDM ensemble 
which underlies these phenomenological successes, and this is an 
inherent property of the DDM framework which transcends the particular calculational procedures
and astrophysical parameters chosen.
Even in the most dramatic case in which we replace the MED propagation model with 
the rather extreme MIN propagation model, 
the very small remaining region of DDM parameter space continues
to exhibit the ``smoking gun'' feature 
in which the positron fraction remains above background out to higher energies
beyond those currently probed. 
Thus, while the specific quantitative results obtained using alternative 
approximations or computational tools may differ somewhat, this relative softness  ---
which we have demonstrated contributes significantly to easing phenomenological 
tensions --- is a real and direct consequence of the 
underlying particle physics of DDM ensembles.

Finally, before concluding, 
we emphasize that the criterion in Eq.~(\ref{eq:GammaConsistencyCondit})
does not represent a constraint which our DDM framework must satisfy,
but rather merely defines a phenomenologically interesting regime of parameter space
within that framework.  Indeed, regions of parameter space in which this criterion is
not satisfied can also in principle yield models which reproduce \mbox{AMS-02}
positron-fraction data and at the same time satisfy other constraints on
dark-matter decays.  In general, such models lead to a more mundane set of
predictions for the positron fraction at high energies, including the possibility of
a relatively sharp downturn similar to that expected in traditional dark-matter models of the
positron excess.  Thus, while the observation of 
a slowly falling or relatively level 
positron excess in future \mbox{AMS-02} data
would strongly favor a decaying DDM ensemble over
other possible dark-matter interpretations, the non-observation of
such a signal would not, in and of itself, 
rule out a DDM ensemble as an explanation of that
excess.


\section{Conclusions\label{sec:conclusions}}


In this paper, we have examined the implications of cosmic-ray data --- and in 
particular, the recent \mbox{AMS-02}  measurement of the positron fraction --- for models
constructed within the Dynamical Dark Matter (DDM) framework.
Because the DDM framework generally includes dark-matter particles  
with lifetimes near the current age of the universe,
present-day cosmic-ray data can be expected to have particular relevance for DDM.
Our primary results can be summarized as follows:  

\begin{itemize}

\item First, we have shown that DDM ensembles provide 
  a viable dark-matter explanation of the existing positron excess.  
  This is true even when the relevant astrophysical and cosmological
  constraints are taken into account.  Indeed, as we have seen, the partitioning 
  of the dark-matter abundance $\OmegaDM$ across an entire ensemble of
  dark-sector fields obeying certain scaling relations provides a natural and 
  well-motivated method of softening the $e^\pm$ injection spectrum and reproducing 
  the existing data.  Moreover, for a DDM ensemble whose constituents decay primarily 
  to $\mu^+\mu^-$, we have found that this softening alone is sufficient to obtain 
  consistency with the positron fraction observed by \mbox{AMS-02} and the combined $e^\pm$ 
  flux observed by FERMI.  

\item Second, we have shown that those DDM scenarios which successfully reproduce
  the observed positron excess generically predict that the positron fraction either 
  levels off or falls gradually at higher energies beyond those currently probed.  
  By contrast, conventional 
  dark-matter scenarios for explaining the positron excess generically predict a very different 
  behavior:  an abrupt downturn in the positron fraction at higher energies.  
  Thus, if we attribute the positron excess to dark-matter physics, we may interpret a
  relatively flat positron excess curve as a ``smoking gun'' of the DDM framework.       

\item Finally, we note that in order to reproduce the positron-fraction curve
  observed by PAMELA and \mbox{AMS-02}, a traditional dark-matter candidate typically 
  must be quite heavy, with a mass  
  $m_\chi \gtrsim 1$~TeV.~  By contrast, we have shown that those DDM ensembles which 
  accurately reproduce the observed positron-fraction curve generically tend to 
  include large numbers of lighter constituent particles $\phi_n$, with masses in the range 
  $300\mathrm{~GeV} \lesssim m_n\lesssim 700\mathrm{~GeV}$.  The presence of such 
  lighter particles playing an active role in the dark sector 
  opens up a broader variety of possibilities 
  for detection using other, complementary probes of the dark sector --- including 
  colliders, direct-detection experiments, {\it etc}\/.   
     
\end{itemize}

Moreover, as we have shown in Sect.~\ref{sec:varyinginputs}, 
these results are largely independent of a variety of input assumptions associated 
with the nature of our DDM ensemble or the modelling of the external 
astrophysical environment.

Despite these results, it is important to stress that our analysis
in this paper has been fundamentally predicated 
on an underlying dark-matter interpretation of the positron excess.
However, there do exist alternative explanations
of these cosmic-ray anomalies.
Indeed, as has been shown in Refs.~\cite{HooperAMS,HooperPulsars}, a contribution to $\Phi_{e^+}$ and 
$\Phi_{e^-}$ from a population of nearby pulsars may provide an alternative explanation 
for the positron excess reported by PAMELA, FERMI, and other cosmic-ray experiments.  
This explanation is a compelling one because it offers an origin for this excess in 
terms of standard astrophysical processes rather than new physics.  Moreover, the pulsar 
explanation is also of particular interest in relation to the cosmic-ray phenomenology of 
DDM ensembles.  Indeed, much like the net contribution from a DDM ensemble, the net 
contribution to $\Phi_{e^+}$ and $\Phi_{e^-}$ from a population of nearby pulsars 
represents the sum over a large number of individual contributions.  It is therefore 
reasonable to expect that many of the same characteristic features which arise in the 
$\Phi_{e^+}$ and $\Phi_{e^-}$ spectra associated with DDM ensembles should also appear 
within the range of possible spectra associated with specific pulsar populations.
Indeed, the shape of the overall positron-fraction curve
associated with a collection 
of pulsars may be difficult to distinguish from that associated with a DDM ensemble. 

However, in general one can measure more than the mere {\it shape}\/
of such a differential flux;  one can also study the {\it directionality}\/ 
associated with its individual angular contributions.
Directionality is especially important in this situation
because it provides a critical method of
distinguishing the case of a DDM ensemble
from that of a population of pulsars.
In the pulsar case,  
an anisotropy is expected in the signal contributions as a function of energy
due to the differing positions of the individual contributing pulsars.  In fact,
if a single nearby pulsar (or a small number thereof) is essentially 
responsible for the observed positron excess, this anisotropy should be observable 
over a wide range of energies $E_{e^\pm}$ 
at the Cherenkov Telescope Array (CTA)~\cite{ProfumoPulsarAnisotropy}.
By contrast, it is reasonable to imagine that the constituent particles within a 
DDM ensemble are distributed in roughly the same manner throughout the galactic 
halo, and thus no significant flux anisotropy is to be expected (other than a slight 
overall flux enhancement in the direction of the galactic center, an enhancement 
which is unlikely to be detected at future experiments).  

Provided that the contributions 
of individual pulsars can be resolved, we conclude that forthcoming data on the anisotropy 
of the observed positron flux may play an important role in differentiating 
between a collection of local pulsars and a DDM ensemble as explanations for the 
observed positron excess. 
However, we note that the propagation of cosmic rays in the local environment is
very complicated, and it may ultimately be very difficult to draw any firm conclusions
in the case that no anistropy is ultimately detected.   These issues are discussed
more fully in Ref.~\cite{noanisotropy}.  
Likewise, we also note that there also exist even more prosaic models
that can potentially explain the positron excess.  These include, for example, models~\cite{secondary} 
in which the excess positrons are generated 
as secondary products of
hadronic interactions inside 
natural cosmic-ray sources such as supernova remnants,
and are thus naturally accelerated in a way 
that endows them with a relatively flat spectrum.
Measurements of the secondary
nuclei produced by cosmic-ray spallation could potentially be used in order
to discriminate between these possibilities~\cite{secondarytwo}.



\begin{acknowledgments}


We would like to thank V.~Bindi, D.~Hooper, I.~Low, D.~Marfatia, P.~Sandick,
J.~Siegal-Gaskins, and X.~Tata for discussions.   KRD is supported in part by the 
Department of Energy under Grants DE-FG02-04ER-41298 and DE-FG02-13ER-41976,
and in part by the National Science Foundation through its employee IR/D program.
JK is supported in part by  DOE Grants DE-FG02-04ER-41291 and DE-FG02-13ER-41913.  
BT is supported in part by DOE Grant DE-FG02-04ER-41291.  The opinions and 
conclusions expressed herein are those of the authors, and do not represent either 
the Department of Energy or the National Science Foundation.

\end{acknowledgments}


\end{document}